\documentclass[reqno,oneside,11pt]{amsart} \oddsidemargin 0.25in
\usepackage{amscd}
\numberwithin{equation}{section}

\newcommand{\states}[1]{\mathsf{#1}}
\newcommand{\reals}[1]{\mathtt{#1}}
\newcommand{\beable}[2]{\alg{B}(#1,#2)}
\newcommand{\cstar}[1]{C^{*}(#1)}
\newcommand{\abs}[1]{\lvert#1\rvert}
\newcommand{\norm}[1]{\lVert#1\rVert}
\newcommand{\ket}[1]{\lvert#1\rangle}
\newcommand{\nullspace}[1]{\mathcal{N}(#1)}
\newcommand{\alg}[1]{\mathfrak{#1}}
\newcommand{\trace}[1]{\mathrm{Tr}(#1)}

\theoremstyle{plain}
\newtheorem{prop}{Proposition}[section]
\newtheorem{lemma}[prop]{Lemma}
\newtheorem{thm}[prop]{Theorem}
\newtheorem{cor}[prop]{Corollary}
\theoremstyle{definition}
\newtheorem*{open}{Open Problem}
\newtheorem*{defn}{Definition}
\newtheorem*{example}{Example}
\theoremstyle{remark}
\newtheorem{remark}[prop]{Remark}
\newtheorem*{notation}{Notation}

\DeclareMathOperator*{\wlim}{WOT-lim}

\newcommand{\clo}[1]{\mathrm{clo}\,\mathtt{#1}}
\newcommand{\spectrum}[1]{\mathrm{sp}(#1)}
\newcommand{\hil}[1]{\mathcal{#1}}
\newcommand{\range}[1]{\mathcal{R}(#1)}
\newcommand{\bh}{\mathfrak{L}(\mathcal{H})}
\newcommand{\vn}[1]{W^{*}(#1)}

\begin{document} \renewcommand{\thefootnote}{\fnsymbol{footnote}}
\begin{center}
{\LARGE Maximal Beable Subalgebras of 
Quantum-Mechanical~Observables\footnote{This work has been submitted 
to Academic Press (which publishes \emph{The Journal of Functional 
Analysis}) for possible publication.  Copyright may be transferred 
without notice, after which this version may no longer be accessible.} \par}%
\vskip 3em%
\lineskip .75em%
{\Large Hans Halvorson\footnote{E-mail: hphst1+@pitt.edu} \par} \vskip
.75em
{\textit{Departments of Mathematics and Philosophy, \\
    University of Pittsburgh, Pittsburgh, PA~15260}
\par} \vskip .75em
{\Large and \par} \vskip .75em {\Large Rob Clifton\footnote{E-mail:
    rclifton+@pitt.edu} \par} \vskip .75em {\textit{Departments of
    Philosophy and History and Philosophy of Science, \\ University of
    Pittsburgh, Pittsburgh, PA~15260} \par}
\vskip 1.5em%
{\large \date{May 12 1999} \par}%
\end{center}
\newpage {\noindent Proposed Running Head: \scshape{Maximal Beable
    Subalgebras}
\par} \vskip 3em
\noindent Please address correspondence to: \begin{tabbing}
\hspace*{1.6in} \= Hans Halvorson \\
\> Department of Philosophy \\
\> University of Pittsburgh \\
\> 1001 Cathedral of Learning \\
\> Pittsburgh, Pennsylvania  15260  \\
\> E-mail: hphst1+@pitt.edu \\
\> Fax: 412-624-5377 \end{tabbing} \newpage
{\begin{center} \scshape{Abstract:} \end{center} \par}
Given a state on an algebra of bounded quantum-mechanical observables, 
we investigate those subalgebras that are maximal with respect to the
property that the given state's restriction to the subalgebra is a
mixture of dispersion-free states---what we call maximal \emph{beable} 
subalgebras (borrowing terminology due to J. S. Bell). 
We also extend our results to the theory of algebras of 
unbounded observables (as developed by Kadison), and show how our results 
articulate a solid mathematical foundation 
for certain tenets of the orthodox `Copenhagen' interpretation of 
quantum theory.  \par \vskip 1.5em
{\begin{center} \scshape{Key words:} \end{center} \par}
Jordan-Lie-Banach Algebra, $C^{*}$-algebra, von Neumann Algebra,
Dispersion-Free State, Pure State, Mixed State \newpage
\section{Introduction}
A number of results in the theory of operator algebras establish the
impossibility of assigning simultaneously determinate values to all
observables of a quantum system.  Von Neumann first observed that the
algebra of bounded operators $\bh$ on a separable Hilbert space with
$\dim\hil{H}>1$ does not admit any dispersion-free normal
state~\cite[p.  320]{von}, a result which was only somewhat later
extended by Misra to arbitrary dispersion-free states \cite[Cor.
2]{mis}.  From general algebraic postulates for observables, and
without recourse to a Hilbert space representation, Segal deduced that
an algebra of quantum-mechanical observables possesses a \emph{full
  set} of dispersion-free states if and only if it is
commutative~\cite[Thm. 3]{seg}.  Kochen and Specker~\cite{ks} relaxed
von Neumann's requirement that values of observables be given by a
linear functional on $\bh$, and regarded the latter as, instead, a
partial algebra, with the product and sum of two elements defined only
if they commute.  For $2<\dim\hil{H}<\infty$, they constructed
finitely generated partial subalgebras of self-adjoint operators in
$\bh$ that possess no \emph{partial} dispersion-free states.  This
result was established independently by J. S. Bell~\cite{bell1}
(cf.~\cite{cliffy}), who also noticed that the nonexistence of a
partial dispersion-free state on \emph{all} self-adjoint elements of
$\bh$ is an immediate corollary of Gleason's theorem~\cite{gleas}.
More recently, these results have been extended to the case of
infinite-dimensional $\hil{H}$, by reduction to the finite case
\cite{jost,bell0}; and, in the latter case, many more examples have
been uncovered of partial subalgebras of observables without partial
dispersion-free states (see~\cite[Ch. 7]{peres},\ \cite[Ch.  3]{bub}
for reviews).

Evidently none of these negative results settle the positive question
of which subalgebras of quantum-mechanical observables (apart from
commutative ones) \emph{can} be taken to have simultaneously
determinate values.  Bell~\cite[Sec.  3]{bell1}, a well-known critic
of the foundational importance of von Neumann's (and, indeed, his own)
result (cf.~\cite[Sec. 3]{cli98}), was the first to raise the
importance of this positive question.  With the aim of avoiding
primitive reference to the term `measurement' in the axiomatic
foundations of quantum theory, Bell forcefully argued (see also
\cite[Chs.  7,19]{bell3}) that it ought to suffice to assign
simultaneous values to some appropriate proper subset of all
quantum-mechanical observables---which he distinguished from the
latter by calling them `\emph{be}ables': ``Could one not just promote
\emph{some} of the observables of the present quantum theory to the
status of beables?  The beables would then be represented by linear
operators in the state space.  The values which they are allowed to
\emph{be} would be the eigenvalues of those operators.  For the
general state the probability of a beable \emph{being} a particular
value would be calculated just as was formerly calculated the
probability of \emph{observing} that value''~\cite[p. 688]{bell2}.
Bell's remarks here suggest the following problem (that has received
scant attention in the mathematical literature; but see
\cite{zimba,cli98}): Given a state on an algebra of observables,
characterize those subalgebras, of `beables,' that are maximal with
respect to the property that the given state's restriction to the
subalgebra is a mixture of dispersion-free states.  Such \emph{maximal
  beable subalgebras} could then represent maximal sets of observables
with simultaneously determinate values distributed in accordance with
the state's expectation values.  The aim of the present paper is to
investigate maximal beable subalgebras and establish their importance
for the foundations of quantum theory.  Later on, we shall extend our
analysis of maximal beable subalgebras to include the case where sets
of \emph{un}bounded observables are assigned simultaneously
determinate values consistent with a state's expectation values.
Though some open problems remain, our results suffice to articulate
certain aspects of the orthodox `Copenhagen' interpretation of quantum
theory (such as the joint indeterminacy of canonically conjugate
variables, and Bohr's defense of the `completeness' of quantum theory
against the argument of Einstein-Podolsky-Rosen) in a mathematically
rigorous way.

\subsection{From JLB- to $C^{*}$-algebras.}
In the first instance, our investigation concerns \emph{algebras} of
bounded quantum-mechanical observables, which immediately raises the
question of what sort of algebraic structure should be assumed.
Following~\cite{lan97,lan,cli98}, we choose to regard the observables
of a quantum system as a JLB-algebra.  In brief, a JLB-algebra is any
real Banach space $(\alg{X},\norm{\cdot},\circ ,\bullet )$ such that
the Jordan product $\circ$ is symmetric, the Lie product $\bullet$ is
antisymmetric and satisfies the Jacobi identity, $\bullet$ is a
derivation with respect to $\circ$, and $\circ$ and $\bullet$ together
respect the associator identity:
\begin{equation} (A\circ B)\circ C-A\circ (B\circ C)=r((A\bullet
  C)\bullet B), \label{assoc}
\end{equation}
for some $r\in [0,\infty )$.  Moreover, defining $A^{2}\equiv A\circ
A$ the norm on $\alg{X}$ must satisfy \begin{equation} \label{norm_sq}
  \norm{A\circ B}\leq \norm{A}\ \norm{B},\ \ \ \norm{A^{2}}
  =\norm{A}^{2},\ \ \ \norm{A^{2}} \leq \norm{A^{2}+B^{2}},
\end{equation} for all $A,B\in \alg{X}$.  

A JLB-algebra $\alg{X}$ has a positive cone $\alg{X}^{+}$ consisting
of elements of the form $\{ A^{2}:A\in \alg{X} \}$.  A linear
functional $\rho$ of $\alg{X}$ is said to be \emph{positive} just in
case $\rho (A)\geq 0$, for all $A\in \alg{X}^{+}$.  If $\alg{X}$ has a
unit $I$, a positive linear functional $\rho$ of $\alg{X}$ is said to
be a state just in case $\rho (I)=1$.

We have not provided any sort of axiomatic or operational derivation
to justify our choice of JLB-algebras over the various other sorts of
algebraic structures we might have used.  For example, we might well
have chosen to set our investigation in the context of Segal algebras
~\cite{seg}, which admit neither a Lie product nor a
\emph{distributive} Jordan product.  However, the choice of
JLB-algebras has an extremely strong pragmatic justification, owing to
the fact that the theory of JLB-algebras (unlike the case of Segal
algebras---see \cite{bobo}) may essentially be reduced to the theory
of $C^{*}$-algebras.

First, if $\alg{X}$ is a JLB-algebra, its complex span
$\alg{X}_{\mathbb{C}}$ is canonically isomorphic to a $C^{*}$-algebra.
In particular, for $A,A'\in \alg{X}$, we define a $C^{*}$ product by
\begin{equation} AA'\equiv (A\circ A')-i\sqrt{r}(A\bullet A'),
\label{cprod} \end{equation}
and, for $A+iB,A'+iB'\in \alg{X}_{\mathbb{C}}$,
\begin{equation} (A+iB)(A'+iB')\equiv (AA'-BB')+i(AB'+BA'). \end{equation}
(The associativity of the $C^{*}$ product follows from the Jacobi and
associator identities together with the fact that $\bullet$ is a
derivation with respect to $\circ$.)  We define an involution $*$ on
$\alg{X}_{\mathbb{C}}$ by
\begin{equation} (A+iB)^{*}\equiv A-iB. \label{cstar} \end{equation}
It can then be shown that the norm on $\alg{X}$ extends uniquely so
that $(\alg{X}_{\mathbb{C}},\norm{\cdot})$ is a $C^{*}$-algebra
(see~\cite[Sec. 3.8]{lan97}).

Conversely, if $(\alg{A},\norm{\cdot})$ is a $C^{*}$-algebra, then the
set of self-adjoint elements of $\alg{A}$, denoted by
$\alg{A}_{\mathrm{sa}}$, forms a real Banach space with norm
$\norm{\cdot}$.  We can then equip $\alg{A}_{\mathrm{sa}}$ with a
Jordan product $\circ$ defined by \begin{equation} A\circ B\equiv
  \frac{1}{2}[A,B]_{+}= \frac{1}{2}(AB+BA), \label{jord}
\end{equation} and with a Lie product
$\bullet$ defined by
\begin{equation} A\bullet B\equiv \frac{i}{2}[A,B]=  \frac{i}{2}(AB-BA).
\label{lie} \end{equation}
The resulting object $(\alg{A}_{\mathrm{sa}},\norm{\cdot},\circ
,\bullet )$ can then be shown to satisfy the axioms which define a
JLB-algebra~\cite{emch,lan97,lan}.
 
Recall that a state of a $C^{*}$-algebra is a positive linear
functional of norm-$1$.  It then follows that there is a natural
bijective correspondence between states of a $C^{*}$-algebra $\alg{A}$
and the states of the JLB-algebra $\alg{A}_{\mathrm{sa}}$.  Indeed, if
$\omega$ is a state of $\alg{A}$, then $\omega
|_{\alg{A}_{\mathrm{sa}}}$ is a state of $\alg{A}_{\mathrm{sa}}$.
Conversely, if $\rho$ is a state of $\alg{A}_{\mathrm{sa}}$, then the
unique linear extension of $\rho$ to $\alg{A}$ is a state of
$\alg{A}$.  (That the extension is indeed a state follows from the
fact that any positive element in a $C^{*}$-algebra is the square of a
self-adjoint element.)

We note two further parallels between JLB- and $C^{*}$-algebras:
\begin{enumerate}
\item A JLB-algebra $\alg{X}$ is called \emph{abelian} just in case
  $A\bullet B=0$ for all $A,B\in \alg{X}$.  Clearly, $\alg{X}$ is
  abelian if and only if $\alg{X}_{\mathbb{C}}$ is an abelian
  $C^{*}$-algebra.
\item Let $\alg{A}$ be a \emph{concrete} $C^{*}$-algebra, acting on
  some Hilbert space $\hil{H}$.  Let $\alg{A}^{-}$ denote the
  weak-operator topology (WOT) closure of $\alg{A}$ in $\bh$.  It then
  follows easily (from the WOT-continuity of $^{*}$ and von Neumann's
  double commutant theorem) that
  $(\alg{A}_{\mathrm{sa}})^{-}=(\alg{A}^{-})_{\mathrm{sa}}$.
  Consequently, $\alg{A}$ is a von Neumann algebra if and only if
  $\alg{A}_{\mathrm{sa}}$ is WOT-closed.
\end{enumerate}

In the remainder of this paper, then, we will carry out our inquiry in
the setting of the theory of $C^{*}$- and von Neumann algebras.  If
the reader is disturbed by the use of complex $^{*}$-algebras in a
discussion of assigning determinate values to quantum-mechanical
\emph{observables} (which, of course, have to be self-adjoint), the
above results can be used to translate what follows into the language
of JLB-algebras.

\subsection{Dispersion-free states.} 
Let $\alg{A}$ be a unital $C^{*}$-algebra and let $\omega$ be a state
of $\alg{A}$.  Following~\cite[p. 304]{and}, we define the
\emph{definite algebra} of $\omega$ by: \begin{equation}
  \alg{D}_{\omega}\equiv\{ A\in \alg{A}:\omega (AX)=\omega (A)\omega
  (X)\;\text{for all}\;X\in \alg{A} \} . \end{equation} It is not
difficult to show that $\alg{D}_{\omega}$ is a unital subalgebra of
$\alg{A}$.  (Indeed (cf.  Exercise 4.6.16 in \cite{kad})
$\alg{D}_{\omega}$ is none other than the complex span of the
Kadison-Singer definite set~\cite[p. 398]{kad59}:
\begin{equation}
\{ A\in
  \alg{A}_{\mathrm{sa}}:\omega (A^{2})=(\omega (A))^{2}\}
  \end{equation}
  which is the JLB-algebra canonically determined, via (\ref{jord})
  and (\ref{lie}), by $\alg{D}_{\omega}$.)  For $A\in \alg{A}$, we say
  that $\omega$ is \emph{dispersion-free} on $A$ just in case $A \in
  \alg{D}_{\omega}$.  If $\alg{B}$ is a subalgebra of $\alg{A}$, we
  say that $\omega$ is \emph{dispersion-free} on $\alg{B}$ just in
  case $\alg{B}\subseteq \alg{D}_{\omega}$.  With this notation, we
  have the following:

\begin{prop} \ \  
\begin{enumerate} 
\item If $\alg{B}\subseteq \alg{D}_{\omega}$, then $\omega
  \,|_{\alg{B}}$ is a pure state.  The converse holds if $\alg{B}$ is
  abelian.
\item If $A\in \alg{D}_{\omega}$, then $\omega (A)\in \spectrum{A}$.
 \end{enumerate} \label{mult_dfree}
\end{prop}

\begin{proof} (i) follows immediately from Proposition~4.4.1
  in~\cite{kad}, and the comments following the proof of that
  proposition.  (ii) follows immediately from Remark~3.2.11
  of~\cite{kad}.
\end{proof}

\begin{remark}  The fact that $\bh$ (with $\hil{H}$ nontrivial and separable) 
  possesses no dispersion-free states can now be easily seen to follow
  from (i) and the fact (cf.~\cite[p. 305]{and}) that
  commutators---i.e., operators expressible as $[X,Y]$ for some
  $X,Y\in\bh$---are norm dense in $\bh$.  Note that if (ii) did not
  hold, it would not make physical sense to use the value of a
  dispersion-free state to represent the intrinsic, possessed value of
  an observable (assuming, that is, that when an observable with a
  determinate value is measured, its value is faithfully revealed by
  the result of the measurement).  \label{herro}\end{remark}

In their partial algebraic approach, Kochen and Specker explicitly
require, not just that (partial) dispersion-free states preserve the
continuous functional relations between observables, but all
\emph{Borel} functional relations~\cite[Eqn. 4]{ks}.  If one restricts
to the case of observables on finite-dimensional spaces (as they
eventually do~\cite[Sec. 3]{ks}), then this extra assumption is
redundant, since every Borel function of such an observable is a
polynomial function.  However, we certainly want to allow as `beables'
observables with continuous spectra, and also assign them (precise
point) values via dispersion-free states.  We end this section by
showing how this allowance forces one to give up requiring that the
values of beables preserve all Borel functional relations.

\begin{defn} If $\alg{V}$ is a von Neumann algebra, and
  $\omega$ is a dispersion-free state of $\alg{V}$, we say that
  $\omega$ satisfies Borel-FUNC on $\alg{V}$ just in case $\omega
  (f(A))=f(\omega (A))$, for each $A\in \alg{V}_{\mathrm{sa}}$ and
  each bounded Borel function $f$ on $\spectrum{A}$.  \end{defn}

Recall, a $*$-homomorphism $\Phi$ from von Neumann algebra
$\alg{V}_{1}$ to von Neumann algebra $\alg{V}_{2}$ is called
\emph{$\sigma$-normal} when $\Phi$ maps the least upper bound of each
increasing sequence of self-adjoint operators bounded above in
$\alg{V}_{1}$ onto the least upper bound of the image sequence in
$\alg{V}_{2}$.  Recall also that a state $\omega$ on a von Neumann
algebra $\alg{V}$ is called \emph{normal} just in case $\omega
(H_{a})\rightarrow \omega (H)$ for each monotone increasing net of
self-adjoint operators $\{ H_{a} \}$ in $\alg{V}$ with least upper
bound $H$.  If $\omega$ is a dispersion-free state of $\alg{V}$, then
$\omega$ (being hermitian) is a $*$-homomorphism of $\alg{V}$ onto the
(von Neumann algebra of) complex numbers.  Thus, a dispersion-free
normal state $\omega$ of $\alg{V}$ is a $\sigma$-normal homomorphism
of $\alg{V}$ onto $\mathbb{C}$.

\begin{notation} If $x$ is a unit vector in $\hil{H}$, $\omega _{x}$ denotes
  the \emph{vector state} of $\bh$ defined by $\omega _{x}(A)=\langle
  Ax,x \rangle$, for each $A\in \bh$.  \end{notation}

\begin{thm} Let $\alg{V}$ be a von Neumann algebra acting on a
  separable Hilbert space $\hil{H}$, and let $\omega$ be a
  dispersion-free state of $\alg{V}$.  Then $\omega$ satisfies
  Borel-FUNC on $\alg{V}$ if and only if there is a unit vector $x\in
  \hil{H}$ such that $\omega =\omega _{x}|_{\alg{V}}$.
  \label{thm_tant}\end{thm}

\begin{proof} ``$\Leftarrow$'' If $\omega =\omega _{x}|_{\alg{V}}$, then 
  $\omega$ is a normal state of $\alg{V}$~(cf.~\cite[Thm.
  7.1.12]{kad}).  Since $\omega$ is (by hypothesis) dispersion-free on
  $\alg{V}$, $\omega$ is a $\sigma$-normal homomorphism of $\alg{V}$
  onto $\mathbb{C}$, and the conclusion that $\omega$ satisfies
  Borel-FUNC on $\alg{V}$ follows immediately from~\cite[Prop.
  5.2.14]{kad}.
  
  ``$\Rightarrow$'' Suppose that $\omega$ satisfies Borel-FUNC on
  $\alg{V}$.  Let $\{ P_{a}:a\in \mathbb{A}\}$ be any family of
  mutually orthogonal projections in $\alg{V}$.  Since $\hil{H}$ is
  separable, $\mathbb{A}$ must be countable, and we may assume that
  $\mathbb{A}=\mathbb{N}$.  Let $A=\sum_{n=1}^{\infty} 3^{-n}P_{n}$.
  Since $\alg{V}$ is SOT-closed, $A\in \alg{V}_{\mathrm{sa}}$.
  Further, $f_{n}(A)=P_{n}$, where $f_{n}$ is the characteristic
  function of the (singleton) set $\{ 3^{-n} \}$.  Let $f$ be the
  characteristic function of the (entire) set $\{ 3^{-n}:n\in
  \mathbb{N} \}$.  Now, $\sum _{n=1}^{\infty} f_{n}=f$ in the sense of
  pointwise convergence of partial sums.  Since the map $g\rightarrow
  g(A)$ from Borel functions on $\spectrum{A}$ into $\alg{V}$ is a
  $\sigma$-normal homomorphism~\cite[p. 320]{kad}, $\sum
  _{n=1}^{\infty} f_{n}(A)=f(A)$.  Using the previous two facts, we
  may compute:
\begin{align}
  \omega (\textstyle \sum P_{n}) &=\:\omega (\textstyle \sum
  f_{n}(A))\:=\:\omega (f(A))\:=\:f(\omega (A)) \\
  &=\:\textstyle \sum f_{n}(\omega (A)) \:=\:\textstyle \sum \omega
  (f_{n}(A)) \:=\:\textstyle \sum \omega (P_{n}), \end{align} where we
used Borel-FUNC in the third and fifth equalities.  Since $\{ P_{n}\}$
is an arbitrary family of orthogonal projections in $\alg{V}$,
$\omega$ is totally additive on $\alg{V}$.  By~\cite[Thm. 7.1.9]{kad},
there is a sequence $\{ x_{n} \}$ of unit vectors in $\hil{H}$, and
sequence $\{ \lambda _{n}\}$ of non-negative real numbers with sum 1,
such that $\omega =\sum _{n=1}^{\infty} \lambda_{n}\omega
_{x_{n}}|_{\alg{V}}$.  We may assume that $0<\lambda _{1}\leq 1$, and
the last equation can then be written as $\omega =\lambda _{1}\omega
_{x_{1}}|_{\alg{V}}+(1-\lambda_{1})\rho$, with $\rho$ a state of
$\alg{V}$.  Since $\omega$ is dispersion-free on $\alg{V}$, it is pure
on $\alg{V}$~(Proposition~\ref{mult_dfree}.(i)).  Thus, $\omega
=\omega _{x_{1}}|_{\alg{V}}$, and $\omega$ is a vector state.
\end{proof}

Evidently a dispersion-free state on a concrete $C^{*}$-algebra is a
vector state only if that vector is a common eigenvector for every
observable in the algebra.  Thus Theorem~\ref{thm_tant} shows
(\emph{pace}~\cite{ks,tel}) that it would be too strong to require
dispersion-free states on beable subalgebras to satisfy Borel-FUNC,
for that would have the effect of excluding continuous spectrum
observables from beable subalgebras by fiat.

\section{Beable and Maximal Beable Subalgebras of Observables}
We start by formalizing the idea of a beable subalgebra, and then
fully characterize both beable and maximal beable subalgebras relative
to a normal (i.e. ultraweakly continuous) state on a concrete
$C^{*}$-algebra (generalizing~\cite[Thm. 10]{cli98}).

\begin{defn} Let $\alg{A}$ be a unital $C^{*}$-algebra, let $\alg{B}$
  be a subalgebra of $\alg{A}$ such that $I\in \alg{B}$, and let
  $\rho$ be a state of $\alg{A}$.  Following~\cite{cli98}, we say that
  $\alg{B}$ is \emph{beable for} $\rho$ if $\rho \,|_{\alg{B}}$ is a
  mixture of dispersion-free states; i.e. if and only if there is a
  probability measure $\mu$ on the space $\states{S}$ of
  dispersion-free states of $\alg{B}$ such that
   \begin{xalignat}{2} \rho (A)= &\int_{\states{S}}\omega _{s}(A)d\mu (s),
     &(A\in \alg{B}). \label{mix}
\end{xalignat}
Physically, $\alg{B}$ is beable for $\rho$ just in case the
observables in $\alg{B}$ can be taken to have determinate values
statistically distributed in accordance with $\rho$'s expectation
values. We say that $\alg{B}$ is \emph{maximal beable} for $\rho$ if
$\alg{B}$ is beable for $\rho$ and $\alg{B}$ is not properly contained
in any other subalgebra of $\alg{A}$ that is beable for $\rho$.  (An
easy application of Zorn's lemma, using the characterization in
Prop.~\ref{beable_equiv} (ii) below, establishes that maximal beable
subalgebras always exist for any state.)
\end{defn}

\begin{example}[Definite Algebra] Let $\alg{A}$ be a $C^{*}$-algebra and let
  $\rho$ be a state of $\alg{A}$.  Clearly $\alg{D}_{\rho}$ is beable
  for $\rho$, since $\rho$ itself is dispersion-free on
  $\alg{D}_{\rho}$.  Although it requires a non-trivial
  result~\cite[Thm. 4]{kad59}, it can also be shown that for any pure
  state $\rho$, $\alg{D}_{\rho}$ is \emph{maximal} beable for
  $\rho$~\cite[Thm. 11]{cli98}.  In the case when $\rho=\omega_{x}$, a
  vector state on a concrete $C^{*}$-algebra, $\alg{D}_{\omega_{x}}$
  consists of exactly those observables with $x$ as an eigenvector.
  For example, Dirac~\cite[Sec. 12]{dirac} takes for granted that the
  observables determinate for a quantum system in a pure state $\omega
  _{x}$ coincide with $\alg{D}_{\omega_{x}}$---an assumption sometimes
  called the `eigenstate-eigenvalue link'.
\label{definite} \end{example}
\begin{notation} If $\hil{M}$ is a subset of some Hilbert space $\hil{H}$, we let $[\hil{M}]$
  denote its closed, linear span.  If $A\in \bh$, we let $\range{A}$
  denote the closure of the range of $A$, and we let $\nullspace{A}$
  denote the null-space of $A$.  If $\hil{T}$ is a closed subspace of
  $\hil{H}$, we let $P_{\hil{T}}$ denote the projection onto
  $\hil{T}$.  For $x\in \hil{H}$, we abbreviate $P_{[x]}$ by $P_{x}$.
\end{notation}

We will make frequent use of the following simple Lemma.

\begin{lemma} Let $\hil{H}$ be a Hilbert space, and let $x\in
  \hil{H}$.  Suppose that $\alg{A}$ is a $C^{*}$-algebra acting on
  $\hil{H}$.  Then, for any $T\in \bh$, if $Tx\in [\alg{A}x]$ and
  $TAx=ATx$ for all $A\in \alg{A}$, then (i) $T$ leaves $[\alg{A}x]$
  invariant, and (ii) $TAy=ATy$ for all $A\in \alg{A}$ and for all
  $y\in [\alg{A}x]$.
  \label{simplify} \end{lemma}

\begin{proof}  (i) Suppose that $Tx\in [\alg{A}x]$ and that $TAx=ATx$ for
  all $A\in \alg{A}$.  Fix $A\in \alg{A}$.  Clearly $A$ itself leaves
  $[\alg{A}x]$ invariant (since $\alg{A}$ is a $C^{*}$-algebra and $A$
  is continuous).  Thus, $TAx=A(Tx)\in [\alg{A}x]$.  Since $A\in
  \alg{A}$ was arbitrary, it follows by the linearity and continuity
  of $T$ that $T$ leaves $[\alg{A}x]$ invariant.
  
  (ii) Let $A\in \alg{A}$, and let $y\in [\alg{A}x]$.  Since $[T,A]$
  is linear and continuous, it is sufficient to show that $[T,A]Bx=0$
  for any $B\in \alg{A}$.  But, this is immediate from the fact that
  $[T,AB]x=0$ and $[T,B]x=0$.
\end{proof}

Let $\alg{A}$ be a $C^{*}$-algebra, and let $\rho$ be a state of
$\alg{A}$.  Recall that the \emph{left-kernel} $\alg{I}_{\rho}$ of
$\rho$ is the set of elements $A\in \alg{A}$ such that $\rho
(A^{*}A)=0$.  We may then formulate the following equivalent
conditions for a subalgebra $\alg{B}$ of $\alg{A}$ to be beable for
$\rho$\,:

\begin{prop} Let $\alg{B}$ be a subalgebra of $\alg{A}$.  Let $\rho$ be
  a state of $\alg{A}$, let $(\pi _{\rho},\hil{H}_{\rho},x_{\rho})$ be
  the GNS representation of $\alg{A}$ induced by the state $\rho$, and
  let $\hil{T}\equiv[\pi _{\rho}(\alg{B})x_{\rho}]\subseteq
  \hil{H}_{\rho}$.  Let $(\phi _{\rho}, \hil{G}_{\rho},v_{\rho})$ be
  the GNS representation of $\alg{B}$ induced by $\rho
  \,|\,_{\alg{B}}$.  Then, the following are equivalent:
  \begin{enumerate} \item $\alg{B}$ is beable for $\rho$. 
  \item $[A,B]\in \alg{I}_\rho$ for all $A,B\in \alg{B}$.
  \item $\phi _{\rho} (\alg{B})$ is abelian.
  \item $\pi_{\rho} (\alg{B})P_{\hil{T}}$ is abelian.
  \end{enumerate} \label{beable_equiv}
\end{prop}
\begin{remark} \cite[Thm. 7]{cli98} contains an alternate proof of 
  (i)\,$\Leftrightarrow$ (ii).  \end{remark}
\begin{proof}  We prove (i)\,$\Rightarrow$
  (ii)\,$\Rightarrow $(iii)\,$\Rightarrow$ (i) and then
  (iii)\,$\Leftrightarrow$ (iv).
  
  ``(i)$\Rightarrow $(ii)'' Suppose that $\alg{B}$ is beable for
  $\rho$.  Then, there is a measure $\mu$ on the set $\states{S}$ of
  dispersion-free states of $\alg{B}$ such that~(\ref{mix}) holds.
  Fix arbitrary $A,B\in \alg{B}$.  Since each $\omega$ in $\states{S}$
  is a $*$-homomorphism of $\alg{B}$ into $\mathbb{C}$, and states are
  hermitian, $\omega ([A,B]^{*}[A,B])=\abs{\omega
    ([A,B])}^{2}=\abs{\omega(A)\omega(B)- \omega(B)\omega(A)}^{2}=0$
  for each $\omega \in \states{S}$, and thus $\rho ([A,B]^{*}[A,B])=0$
  by~(\ref{mix}).
  
  ``(ii)$\Rightarrow $(iii)'' Suppose that $[A,B]\in \alg{I}_{\rho}$
  for all $A,B\in \alg{B}$.  In order to show that $\phi
  _{\rho}(\alg{B})$ is abelian, let $\phi _{\rho}(A)\in \phi
  _{\rho}(\alg{B})$.  Thus, for any $\phi _{\rho}(B)\in \phi
  _{\rho}(\alg{B})$,
  \begin{gather} \bigg\langle \Big[\phi _{\rho}(A),\phi 
    _{\rho}(B)\Big]v_{\rho}\,,\, \Big[\phi _{\rho}(A),\phi _{\rho}
    (B)\Big]v_{\rho}\bigg\rangle \: =\:\bigg\langle \phi
    _{\rho}\Big([A,B]^{*}[A,B]\Big)v_{\rho}\,,\,v_{\rho}
    \bigg\rangle \\
    =\rho \Big([A,B]^{*}[A,B]\Big)\:=\: 0 . \end{gather} Thus, $[\phi
  _{\rho}(A),\phi _{\rho}(B)]v_{\rho}=0$.  Now, since $[\phi
  _{\rho}(\alg{B})v_{\rho}]=\hil{G}_{\rho}$ we may apply
  Lemma~\ref{simplify}, with $C^{*}$-algebra $\phi _{\rho} (\alg{B})$
  and vector $v_{\rho}$, to conclude that $\phi _{\rho}(A)\in \phi
  _{\rho}(\alg{B})'$.  Therefore, $\phi _{\rho}(\alg{B})\subseteq \phi
  _{\rho}(\alg{B})'$, and $\phi _{\rho}(\alg{B})$ is abelian.
  
  ``(iii)$\Rightarrow $(i)'' If $\phi _{\rho}(\alg{B})$ is abelian, we
  may identify it with the set of continuous, complex-valued
  functions, $\mathcal{C}(\states{S})$, on some compact Hausdorff
  space $\states{S} $~\cite[Thm.  4.4.3]{kad}.  Consider the vector
  state $\omega _{v_{\rho}}$ on $\phi
  _{\rho}(\alg{B})\simeq\mathcal{C}(\states{S} )$ induced by
  $v_{\rho}$.  By the Riesz Representation Theorem~\cite[Thm.
  2.14]{rud}, there is a probability measure $\mu$ on $\states{S}$
  such that
\begin{xalignat}{2} \omega _{v_{\rho}}(\phi (A)) &=\int _{S} [\phi _{\rho}
  (A)](s)d\mu (s), &(\phi _{\rho}(A) \in \phi _{\rho}
  (\alg{B})\simeq\mathcal{C}(\states{S})). \end{xalignat} For each
$s\in \states{S}$, define $\omega _{s}:\alg{B}\rightarrow \mathbb{C}$
by $\omega _{s}(A) =[\phi _{\rho}(A)](s),\;(A\in \alg{B})$.  The
reader may verify without difficulty that each $\omega _{s}$ defines a
dispersion-free state on $\alg{B}$ (using the fact that $\phi _{\rho}$
is a $*$-homomorphism, and the definition of multiplication on
$\mathcal{C}(\states{S})$).  Finally, for each $A\in \alg{B}$,
\begin{equation} \begin{split}
    \rho (A)&\equiv (\omega _{v_{\rho}}\circ \phi
    _{\rho})(A)\:=\:\omega
    _{v_{\rho}}(\phi _{\rho}(A)) \\
    &=\int _{\states{S} }[\phi _{\rho}(A)](s)d\mu (s)\:=\:\int
    _{\states{S} }\omega _{s}(A)d\mu (s).
\end{split} \end{equation} Therefore, $\alg{B}$ is beable for $\rho$.

``(iii)$\Leftrightarrow $(iv)'' $\pi_{\rho}(\alg{B})$ is a
$C^{*}$-subalgebra of $\pi_{\rho}(\alg{A})$, which is in turn a
$C^{*}$-subalgebra of $\alg{L}(\hil{H}_{\rho})$.  Thus, the mapping
$\pi _{\rho}(A)\stackrel{\xi}{\rightarrow} \pi_{\rho}(A)|_{\hil{T}}$
is a representation of $\pi_{\rho}(\alg{B})$ on $\hil{T}$ with cyclic
vector $x_{\rho}$~(cf.~\cite[p. 276]{kad}), and the composition map
$\widehat{\pi}_{\rho} \equiv \xi \circ \pi_{\rho}|_{\alg{B}}$ is a
cyclic representation of $\alg{B}$ on $\hil{T}$.

We now show that $(\widehat{\pi}_{\rho},\hil{T})$ is unitarily
equivalent to $(\phi _{\rho},\hil{G}_{\rho})$.  Recall that
$\hil{G}_{\rho}$ is the completion of the pre-hilbert space $\{
A+\alg{I}^{\alg{B}}_{\rho}:A\in \alg{B}\}$, where
$\alg{I}^{\alg{B}}_{\rho}\equiv \alg{I}_{\rho}\cap \alg{B}$.  For
elements of this latter set, there is a natural isometric mapping
$\overline{U}$ into $\hil{T}$; namely the mapping that takes
$A+\alg{I}^{\alg{B}}_{\rho}$ to $A+\alg{I}_{\rho}$.  It is not
difficult to verify that $\overline{U}$ extends uniquely to a unitary
operator $U$ from $\hil{G}_{\rho}$ onto $\hil{T}$, and that
$\widehat{\pi}_{\rho} (A)U=U\phi _{\rho}(A)$ for all $A$ in $\alg{B}$.
Thus, $(\widehat{\pi}_{\rho},\hil{T})$ is unitarily equivalent to
$(\phi _{\rho},\hil{G}_{\rho})$.

The equivalence of (iii) and (iv) now follows from the fact that $\pi
_{\rho}(\alg{B})P_{\hil{T}}$ is $*$-isomorphic to
$\widehat{\pi}_{\rho}(\alg{B})$.  \end{proof}

Recall that a state $\rho$ on a $C^{*}$-algebra $\alg{B}$ is called
\emph{faithful} just in case whenever $A\in \alg{B}^{+}$ and $\rho
(A)=0$, then $A=0$.

\begin{cor} Suppose that $\alg{B}$ is beable for $\rho$ and that $\rho$ is
  a faithful state of $\alg{B}$.  Then $\alg{B}$ is abelian.
  \label{faithful} 
\end{cor}

\begin{proof} Since $\rho$ is faithful, $\phi _{\rho}$ is an isomorphism of
  $\alg{B}$ onto $\phi _{\rho} (\alg{B})$~\cite[Exercise 4.6.15]{kad}.
  However, $\phi _{\rho} (\alg{B})$ is
  abelian~(Prop.~\ref{beable_equiv} (iii)).  \end{proof}

\begin{example}[Vacuum State]  Let $\{ \alg{A}(O) \} _{O\subseteq M}$
  be a net of local von Neumann algebras over Minkowski spacetime $M$,
  and let $\rho$ be the vacuum state~(cf.~\cite[p.  23]{hor}).  If $O$
  has nonempty spacelike complement in $M$, it follows by the
  Reeh-Schlieder Theorem~\cite[Thm. 1.3.1]{hor} that $\rho$ is a
  faithful state of $\alg{A}(O)$ (since $\rho$ is induced by the
  vacuum vector $\Omega$ which is separating for $\alg{A}(O)$).
  Suppose that $\alg{B}\subseteq \alg{A}(O)$ and that $\alg{B}$ is
  beable for $\rho$.  Then $\rho\,|\,_{\alg{B}}$ is faithful, and it
  follows from Corollary~\ref{faithful} that $\alg{B}$ is abelian.
\end{example}

\subsection{Beable Algebras for Normal States.}
We have defined the beable status of a $C^{*}$-algebra $\alg{B}$ with
respect to an arbitrary state $\rho$ of $\alg{B}$.  In what follows,
we specialize to the concrete case where $\alg{B}$ is acting on some
(fixed) Hilbert space $\hil{H}$ (not necessarily separable).  If
$\rho$ is a normal state of $\bh$, it follows that there is a positive
trace-$1$ operator $K\in \bh$ such that $\rho (A)=\trace{KA}$ for each
$A\in \bh$~\cite[Remark 7.1.10, Thm. 7.1.12]{kad}.  With this in mind,
we will freely interchange ``$\alg{B}$ is beable for $\rho$,'' with
``$\alg{B}$ is beable for $K$.''

\begin{notation} In what follows, we will abbreviate $\range{K}$ by
  $\hil{K}$.  \end{notation}

\begin{remark} In the special case where $K=P_{v}$ for some unit vector
  $v\in \hil{H}$, $\alg{B}$ is beable for $P_{v}$ just in case
  $ABv=BAv$ for each $A,B\in \alg{B}$.  This follows by
  Proposition~\ref{beable_equiv} (ii) since
  $\trace{P_{v}[A,B]^{*}[A,B]}=\langle [A,B]v,[A,B]v\rangle$.
  \label{rem_vec} \end{remark}

\begin{lemma} Suppose that $\alg{B}$ is a subalgebra of $\bh$, $K$ is a
  positive, trace-$1$ operator on $\hil{H}$, $\hil{M}$ is a subset of
  $\hil{H}$, and $0\neq v\in \hil{H}$.  \begin{enumerate}
  \item $\alg{B}$ is beable for $P _{x}$, for all $x\in \hil{M}$, if
    and only if $\alg{B}$ is beable for $P_{y}$, for all $y\in
    [\hil{M}]$.
  \item $\alg{B}$ is beable for $P_{v}$ if and only if $\alg{B}$ is
    beable for $P_{x}$, for all $x\in [\alg{B}v]$.
  \item $\alg{B}$ is beable for $K$ if and only if $\alg{B}$ is beable
    for $P_{x}$, for all $x\in \hil{K}$.  \end{enumerate}
  \label{lemma_def} \end{lemma}

\begin{proof} (i) The ``if'' implication is trivial.  Suppose then that
  $\alg{B}$ is beable for $P_{x}$, for all $x\in \hil{M}$.  Consider
  the closed subspace of $\hil{H}$ given by
\begin{equation} \hil{Y}
\equiv \bigwedge \left\{ \nullspace{[A,B]}\;:\;  A,B\in
\alg{B}\right\}.  \label{beable_space}
\end{equation}
Clearly, $\hil{Y}$ is precisely the set of all $x\in \hil{H}$ such
that $\alg{B}$ is beable for $P_{x}$~(see Remark~\ref{rem_vec}).  By
supposition, $\hil{M} \subseteq \hil{Y}$; thus, $\hil{Y}$ will also
contain $\hil{M}$'s closed, linear span $[\hil{M}]$.

(ii) The ``if'' implication is trivial, since $\alg{B}$ contains the
identity.  Conversely, suppose that $\alg{B}$ is beable for $P _{v}$.
Fix $A\in \alg{B}$.  Then, for any $B\in \alg{B}$, $[A,B]v=0$, and
moreover $Av\in [\alg{B}v]$.  Thus, we may apply Lemma~\ref{simplify}
to conclude that $ABx=BAx$ for any $B\in \alg{B}$ and for any $x\in
[\alg{B}v]$.  Since $A$ was an arbitrary element of $\alg{B}$, it
follows (Remark \ref{rem_vec}) that $\alg{B}$ is beable for $P_{x}$
whenever $x\in [\alg{B}v]$.

(iii) Recall, first, that as a positive, trace-$1$ operator, $K$ has a
pure-point spectrum~\cite[pp.  188-191]{von}.  By the spectral theorem
(and the fact that $K$ leaves $\hil{K}$ invariant), $\hil{K}$ is the
closed span of the eigenvectors of $K$ in its range.  Thus, there is a
countable set $\{ x_{n} \} \subseteq \hil{K}$, such that
$\norm{x_{n}}=1$ for all $n$, $K=\sum_{n}\lambda _{n} P_{x_{n}}$,
where $\lambda _{n} \in (0,1]$, and $\sum _{n} \lambda _{n}=1$.

``$\Rightarrow$'' Suppose that $\alg{B}$ is beable for $K$.  Recall
from Proposition~\ref{beable_equiv} (ii) that $\alg{B}$ is beable for
$K$ if and only if $\trace{K[A,B]^{*}[A,B]} =0$ for all $A,B\in
\alg{B}$.  Given any eigenvector $y$ in $\hil{K}$, we may write
$K=\lambda P_{y}+(1-\lambda )K'$ for some positive, trace-$1$ operator
$K'$, and $\lambda \in (0,1]$. Thus, by the linearity of the trace,
\begin{align}
  \lambda \trace{P_{y}[A,B]^{*}[A,B]}
  &=\trace{K[A,B]^{*}[A,B]}-(1-\lambda
  )\trace{K'[A,B]^{*}[A,B]} \\
  &\leq \trace{K[A,B]^{*}[A,B]}=0, \label{tozero}
\end{align}
where the inequality in~(\ref{tozero}) follows since $\lambda \in
(0,1]$ and $\trace{K'[A,B]^{*}[A,B]}\geq 0$.  Thus,
$\trace{P_{y}[A,B]^{*}[A,B]}=0$ for any eigenvector $y$ of $K$ in its
range.  Since the closed linear span of these eigenvectors is just
$\hil{K}$, the conclusion follows by (i).

``$\Leftarrow$'' Let $A,B\in \alg{B}$.  Then, by hypothesis,
$\trace{P_{x}[A,B]^{*}[A,B]}=0$ whenever $x\in \hil{K}$.  In
particular, $\trace{P_{x_{n}}[A,B]^{*}[A,B]}$, for each $n$, where
$K=\sum _{n} \lambda _{n}P_{x_{n}}$.  Therefore,
$\trace{K[A,B]^{*}[A,B]}=\sum _{n} \lambda _{n}
\trace{P_{x_{n}}[A,B]^{*}[A,B]} =0$.  Since $A,B\in \alg{B}$ were
arbitrary, $\alg{B}$ is beable for $K$.
\end{proof}

\begin{lemma} $\alg{B}$ is beable for $K$ if and only if $\alg{B}$ is
  beable for $P_{x}$, for all $x\in [\alg{B}\hil{K}]$.
  \label{cor_def} \end{lemma}

\begin{proof} The ``if'' implication follows trivially from
  Lemma~\ref{lemma_def}.(iii).  Conversely, suppose $\alg{B}$ is
  beable for $K$.  By (iii), $\alg{B}$ is beable for $P_{y}$, for all
  $y\in \hil{K}$.  Fix $y$.  By (ii), $\alg{B}$ is beable for $P_{z}$,
  for all $z\in [\alg{B}y]$.  Finally, $[\alg{B}\hil{K}]=\bigvee
  _{y\in \hil{K}} [\alg{B}y]$, so by (i), $\alg{B}$ is beable for
  $P_{x}$, for all $x\in [\alg{B}\hil{K}]$.  \end{proof}

We turn now to providing intrinsic operator algebraic
characterizations of beable, and maximal beable, status with respect
to a normal state.

\begin{thm} Let $\alg{B}$ be a $C^{*}$-algebra acting on $\hil{H}$, and let
  $\hil{T} \equiv [\alg{B}\hil{K}]$.  Then, \begin{enumerate}
  \item $\alg{B}$ is beable for $K$ if and only if $\alg{B}\subseteq
    \alg{L}(\hil{T}^{\perp})\oplus \alg{N}$, where $\alg{N}$ is an
    abelian subalgebra of $\alg{L}(\hil{T})$.
  \item $\alg{B}$ is maximal beable for $K$ if and only if
    $\alg{B}=\alg{L}(\hil{T}^{\perp})\oplus \alg{N}$, where $\alg{N}$
    is a maximal abelian subalgebra of $\alg{L}(\hil{T})$.
  \end{enumerate}
\label{thm_def} \end{thm}

\begin{proof} (i) ``$\Rightarrow$''  Suppose that $\alg{B}$ is beable for
  $K$.  Clearly, we have defined $\hil{T}$ in such a way that
  $\hil{T}$ reduces $\alg{B}$.  Thus, each element of $\alg{B}$ will
  decompose uniquely into the direct sum of an operator on $\hil{T}
  ^{\perp}$ and an operator on $\hil{T}$.  We must show that whenever
  $A_{1} \oplus A_{2}$ and $B_{1} \oplus B_{2}$ are in $\alg{B}$, then
  $A_{2} B_{2} =B_{2} A_{2}$.  In other words, we must show that
  elements of the set $P_{\hil{T}}\alg{B} P_{\hil{T}}$ commute with
  each other.
  
  Let $A,B\in P_{\hil{T}}\alg{B} P_{\hil{T}}$.  Thus,
  $A=P_{\hil{T}}A'P_{\hil{T}}$ for some $A'\in \alg{B}$ and
  $B=P_{\hil{T}}B'P_{\hil{T}}$ for some $B'\in \alg{B}$.  Let $x\in
  \hil{H}$ be arbitrary.  Then, $x=y+z$ for (unique) $y\in \hil{T}
  ^{\perp }$ and $z\in \hil{T}$.  Since $z\in \hil{T}$, $\alg{B}$ is
  beable for $P_{z}$~(Lemma~\ref{cor_def}).  Thus,
\begin{align}
  ABx &=(P_{\hil{T}}A'P_{\hil{T}}P_{\hil{T}}B')P_{\hil{T}}x\:=\:(P_{\hil{T}}A'P_{\hil{T}}P_{\hil{T}}B')z \\
  &=(P_{\hil{T}}A'P_{\hil{T}})B'z\:=\:(P_{\hil{T}}A')B'z\:=\:A'B'z,
\end{align} where the last two equalities hold because both $A'$ and 
$B'$ leave $\hil{T}$ invariant.  By symmetry, $BAx=B'A'z$.  But,
$A'B'z=B'A'z$ since $A' \, ,B' \in \alg{B}$, $z\in \hil{T}$, and
$\alg{B}$ is beable for $P_{z}$.  Thus, $ABx=BAx$, and since $x$ was
arbitrary, $AB=BA$.  Since $A,B\in P_{\hil{T}}\alg{B}P_{\hil{T}}$ were
arbitrary, any two elements of $P_{\hil{T}}\alg{B} P_{\hil{T}}$
commute.

``$\Leftarrow$'' Suppose that $P_{\hil{T}}\alg{B} P_{\hil{T}}$
consists of mutually commuting operators.  Let $x\in \hil{K}$.  Then,
since $\alg{B}$ contains the identity, $x\in \hil{T}$.  Let $A,B\in
\alg{B}$.  Then, we may write $A=A_{1}\oplus A_{2}$ and $B=B_{1}\oplus
B_{2}$.  Hence, $ABx=(A_{1}\oplus A_{2})(B_{1} \oplus B_{2})x
=(A_{1}\oplus A_{2} )(0+B_{2}x)=A_{2}B_{2}x$.  By symmetry,
$BAx=B_{2}A_{2}x$. But, since elements of
$P_{\hil{T}}\alg{B}P_{\hil{T}}$ commute, $A_{2}B_{2}x=B_{2}A_{2}x$.
Thus, $ABx=BAx$ for any $A,B\in \alg{B}$; that is, $\alg{B}$ is beable
for $P _{x}$.  Furthermore, since $x$ was an arbitrary element of
$\hil{K}$, we see that $\alg{B}$ is beable for every state defined by
a (unit) vector in $\hil{K}$.  By Lemma~\ref{lemma_def}.(iii),
$\alg{B}$ is beable for $K$.

(ii) We have proved in (i) that any algebra $\alg{B}$ which is beable
for $K$ will be commutative in its action on $[\alg{B}\hil{K}]$, and
that any algebra $\alg{B}$ which is commutative in its action on
$[\alg{B}\hil{K}]$ will be beable for $K$.  To complete the proof,
then, it will suffice to show that if
$\alg{B}=\alg{L}(\hil{T}^{\perp})\oplus \alg{N}$, where $\alg{N}$ is
maximal abelian, then $\alg{B}$ is not properly contained in any
beable algebra for $K$.

Suppose then that $\alg{B} \subseteq \alg{C}$, and that $\alg{C}$ is
beable for $K$. (We show that $\alg{C} = \alg{B}$.)  Since $\alg{C}$
is beable for $K$, $\alg{C}$ is beable for $P_{y}$ whenever $y\in [
\alg{C} \hil{K} ]$~(Lemma~\ref{cor_def}).  Furthermore, $[\alg{B}
\hil{K}] \subseteq [ \alg{C} \hil{K} ] $.  Thus, $\alg{C}$ is beable
for $P_{z}$ whenever $z\in [\alg{B} \hil{K}]\equiv \hil{T}$.  Now,
$P_{\hil{T}}\in \alg{N}$ since the latter is maximal abelian and since
the former is the identity on $\hil{T}$.  Thus, $P_{\hil{T}}\in
\alg{B} \subseteq \alg{C}$.  Let $D$ be a self-adjoint element of
$\alg{C}$.  Then, for all $z\in \hil{T}$, $Dz =DP_{\hil{T}}z
=P_{\hil{T}}Dz$, since $\alg{C}$ is beable for $P_{z}$.  That is, $D$
leaves $\hil{T}$ invariant.  However, since $D$ is self-adjoint, it
also leaves $\hil{T}^{\perp}$ invariant, and therefore
$D=(I-P_{\hil{T}})D(I-P_{\hil{T}})\oplus P_{\hil{T}}DP_{\hil{T}} \in
\alg{L}(\hil{T}^{\perp})\oplus \alg{L}(\hil{T})$.

Finally, let $A\in \alg{N}$.  For any $z \in \hil{T}$,
\begin{align}
  P_{\hil{T}}DP_{\hil{T}}Az &=P_{\hil{T}}DAz\:=DAz\:=\:ADz \\
  &=ADP_{\hil{T}}z\:=\:AP_{\hil{T}}DP_{\hil{T}}z. \end{align} The
first, second, and fifth equalities hold since $A$ and $D$ leave
$\hil{T}$ invariant.  The third equality holds since $A,D\in \alg{C}$,
and $\alg{C}$ is beable for $P_{z}$.  Hence,
$P_{\hil{T}}DP_{\hil{T}}\in \alg{N}' \subseteq \alg{N}$ and thus
$D\;\in\;\alg{L}(\hil{T}^{\perp})\oplus \alg{N}$.  We have shown that
$\alg{C}_{\mathrm{sa}} \subseteq \alg{B}$, from which it follows that
$\alg{C}\subseteq \alg{B}$ and $\alg{B}$ is maximal beable for $K$.
\end{proof}

\begin{example}[Multiplication Algebra]  Let $\alg{M}$ be the von
  Neumann algebra of multiplications by essentially bounded
  (measurable) functions on $L_{2}(\mathbb{R})$, generated by the
  unbounded `multiplication by $x$' (position) operator.  Let $\psi$
  be any (wave) function in $L_{2}$ that is non-zero almost
  everywhere.  It follows then that $\alg{M}$ is maximal beable for
  $\psi$.  Indeed, an elementary measure-theoretic argument proves
  that $\psi$ is a separating vector for $\alg{M}$.  Moreover, since
  $\alg{M}$ is maximal abelian, $\alg{M}=\alg{M}'$ and $\psi$ is a
  generating vector for $\alg{M}$~\cite[Cor. 5.5.12]{kad}.  Thus,
  $\hil{T}\equiv [\alg{M}\psi]=L_{2}$ and the maximal beable status of
  $\alg{M}$ for $\psi$ follows from Theorem~\ref{thm_def} (ii).
  Bohm's `causal' interpretation of quantum theory~\cite{bohm}---which
  only grants beable status to a particle's position---can be
  understood as privileging $\alg{M}$ (see~\cite[Sec. 5]{cli98}).
\end{example}

\begin{cor} Let $\rho$ be a normal state on $\bh$.  \begin{enumerate} 
  \item If $\alg{B}$ is maximal beable for $\rho$, then
    $\alg{B}=\alg{B}^{-}$.
  \item If $\alg{B}$ is beable for $\rho$, then $\alg{B}^{-}$ is
    beable for $\rho$ as well.  \end{enumerate}
\label{cor_woclosed}  \end{cor}
\begin{proof} (i) Let $K$ be a positive trace-$1$ operator that induces the
  state $\rho$ on $\bh$.  If $\alg{B}$ is maximal beable for $K$, then
  $\alg{B}=\alg{L}(\hil{T}^{\perp})\oplus \alg{N}$, where $\alg{N}$ is
  a maximal abelian subalgebra of $\alg{L}(\hil{T})$.  Since $\alg{N}$
  is a maximal abelian subalgebra of $\alg{L}(\hil{T})$, it follows
  that $\alg{N}$ is a von Neumann algebra.  Therefore, $\alg{B}$ is a
  von Neumann algebra.
  
  (ii) Now suppose that $\alg{B}$ is beable for $\rho$.  Then,
  $\alg{B}$ is contained in some maximal beable algebra $\alg{C}$ for
  $\rho$.  By part (i) of this Corollary, $\alg{C}=\alg{C}^{-}$.
  Thus, $\alg{B}^{-}\subseteq \alg{C}^{-}=\alg{C}$, and since beable
  status is hereditary, the conclusion follows.  \end{proof}

Recall that a pure state $\rho$ on a concrete $C^{*}$-algebra
$\alg{A}$ is called \emph{singular} just in case it is \emph{not}
ultraweakly continuous.  Thus, a singular state is a pure, non-normal
state.

\begin{remark} Both parts of the above Corollary, in particular (i),
  fail if $\rho$ is not assumed to be a normal state of $\bh$.  For
  example, if $\rho$ is a singular state of $\bh$, then
  $\alg{D}_{\rho}$ is maximal beable for $\rho$~(see
  Example~\ref{definite}).  However, $\alg{D}_{\rho}$ is not
  WOT-closed.  For recall that $\rho \,|_{\alg{K}}=0$, where $\alg{K}$
  is the ideal of compact operators in $\bh$~\cite[Cor. 10.4.4]{kad}.
  Thus, $\alg{K}\subseteq \alg{D}_{\rho}$, since $\rho
  (AX)\!=\!0=\!\rho (A)\rho (X)$ for any $A\in \alg{K}$ and for any
  $X\in \bh$.  Moreover, $\alg{K}^{-}=\bh$, and it follows that
  $(\alg{D}_{\rho})^{-}=\bh $.  But, clearly, $\alg{D}_{\rho}\neq \bh
  $ ($\hil{H}$ separable), since there are no states dispersion-free
  on all of $\bh$~(Remark~\ref{herro}).\end{remark}

\section{Beable Status for Unbounded Observables}
To this point, we have restricted discussion of ``beable status'' to
bounded operators.  Of course, many of the observables of interest in
quantum theory, such as position and momentum, are represented by
unbounded operators.  Thus, in this section we make use of the theory
of algebras of unbounded functions and operators (as expounded
in~\cite[Sec. 5.6]{kad} and Kadison~\cite{kad86}) in order to
articulate the sense in which an unbounded operator can have beable
status with respect to a state.  The section ends with results that
capture the essential content of the Heisenberg-Bohr indeterminacy
principle for canonically conjugate observables.

Let $\alg{V}$ be a von Neumann algebra acting on $\hil{H}$ and let $R$
be a (possibly unbounded) normal operator on $\hil{H}$.  $R$ is said
to be \emph{affiliated} with $\alg{V}$ just in case $U^{*}RU=R$
whenever $U$ is a unitary operator in $\alg{V}'$.  Frequently this
relation is denoted by $R\,\eta\,\alg{V}$.  Now, if $\alg{V}$ is an
\emph{abelian} von Neumann algebra, the set $\states{S}$ of
\emph{pure} states of $\alg{V}$, with the weak-$*$ (i.e. pointwise
convergence) topology, is an extremely disconnected compact Hausdorff
space, and $\alg{V}$ is $*$-isomorphic to
$\mathcal{C}(\states{S})$~\cite[Thm.  4.4.3, Thm.  5.2.1]{kad}.  Under
this isomorphism, $A\in \alg{V}$ goes to $\phi (A)\in
\mathcal{C}(\states{S})$ defined by $\phi (A)(\omega )=\omega (A)$,
for all $\omega \in \states{S}$.  A \emph{normal function} on an
extremely disconnected compact Hausdorff space $\states{S}$ is defined
as a continuous complex-valued function $f$ defined on an open dense
subset $\states{S} \backslash\states{Z}$ of $\states{S}$ such that
$\lim_{\omega\rightarrow \tau }\abs{f(\omega)}=\infty$ for each $\tau$
in $\states{Z}$ (where $\omega\in \states{S} \backslash\states{Z}$),
and a \emph{self-adjoint function} on $\states{S}$ is a real-valued
normal function on $\states{S}$~\cite[Def.  5.6.5]{kad}.  Let
$\mathcal{N}(\alg{V})$ be the set of (normal) operators affiliated
with $\alg{V}$.  Then, $\mathcal{N}(\alg{V})$ may be equipped with two
operations $\widehat{+}$ (closed addition) and $\widehat{\cdot}$
(closed multiplication) under which it is a commutative
$*$-algebra~\cite[Thm.  5.6.15]{kad}.  Similarly, if
$\mathcal{N}(\states{S})$ is the set of normal functions on
$\states{S}$, then there are operations
$\widehat{+}\,,\,\widehat{\cdot}$ and $*$ that extend the standard
pointwise operations.  Moreover, the $*$-isomorphism $\phi$ from
$\alg{V}$ onto $C(\states{S})$ extends to a $*$-isomorphism (which we
denote again by $\phi$) from $\mathcal{N}(\alg{V})$ onto
$\mathcal{N}(\states{S})$, providing us with what we might call the
``extended function representation'' of the abelian von Neumann
algebra $\alg{V}$~\cite[Thm.  5.6.19]{kad}.

For each family $\alg{F}$ of normal operators, there will be a unique
smallest (not necessarily abelian) von Neumann algebra $\vn{\alg{F}}$
such that $R$ is affiliated with $\vn{\alg{F}}$ for all $R\in
\alg{F}$.  We may call $\vn{\alg{F}}$ the von Neumann algebra
generated by $\alg{F}$.  If $\alg{F}$ consists of a single normal
operator $R$, then it follows that $\vn{R}$ is an \emph{abelian} von
Neumann algebra~\cite[Thm. 5.6.18]{kad}.  Thus, $R$ is represented by
a normal function $\phi (R)$ on $\states{S}$, where $\states{S}$ is
now the set of pure states of $\vn{R}$.  If, as usual, $\spectrum{R}$
is defined to be the set of real numbers $\lambda$ such that
$R-\lambda I$ is not a one-to-one mapping of the domain of $R$ onto
$\hil{H}$, it follows that the range of the function $\phi (R)$ is
identical to $\spectrum{R}$~\cite[Proposition 5.6.20]{kad}.  It is not
difficult to see that the range of a normal function is a closed
(compact only if $R$ is bounded) subset of $\mathbb{C}$~\cite[p.
356]{kad}.  Thus, $\spectrum{R}$ is closed in $\mathbb{C}$.

Borel functions of $R$ may be defined, via the isomorphism of
$\mathcal{N}(\vn{R})$ and $\mathcal{N}(\states{S})$, as
follows~\cite[Remark 5.6.25]{kad}.  Let $\states{Z}$ be the closed
nowhere dense subset of $\states{S}$ such that $\phi (R)$ is defined
and continuous on $\states{S} \backslash\states{Z}$.  Let $g$ be an
arbitrary element of $\mathcal{B}_{u}(\spectrum{R})$, the algebra of
complex-valued Borel functions (finite almost everywhere) on
$\spectrum{R}$.  Define $\widetilde{g}$ by:
\begin{equation}
\widetilde{g}(\omega)\equiv
\begin{cases} (g\circ \phi (R))(\omega) &\omega\in \states{S} \backslash\states{Z},  \\
  0 &\omega \in \states{Z}. \end{cases} \label{func_calc}
\end{equation} Then $\widetilde{g}$ is in
$\mathcal{B}_{u}(\states{S})$, and there is a \emph{unique} function
$h\in \mathcal{N}(\states{S})$ such that $\widetilde{g}$ and $h$ agree
on the complement of a meager (i.e. first category) set
$\states{M}$~\cite[Lemma 5.6.22]{kad}.  Note that since $\states{S}$
is compact Hausdorff, the Baire Category Theorem ensures us that
$\states{S} \backslash\states{M}$ is dense in $\states{S}$.  Thus,
$\widetilde{g}$ and $h$ may not disagree on any non-empty open set---a
fact we shall make frequent use of in what follows. Finally, $g(R)$ is
defined as $\phi ^{-1}(h)$, as represented in the diagram below:
\begin{equation*}
\begin{CD}
  \mathcal{B}_{u}(\spectrum{R}) @>{g\rightarrow \widetilde{g}}>>
  \mathcal{B}_{u}(\states{S} ) \\
  @V{g\rightarrow g(R)}VV  @VV{\widetilde{g}\rightarrow h}V  \\
  \mathcal{N}(\vn{R}) @>>{\phi }> \mathcal{N}(\states{S} )
\end{CD}
\end{equation*}
The Borel functional calculus also provides a method of defining a
projection-valued measure $E$ on $\spectrum{R}$ and, by extension, a
projection-valued measure on $\mathbb{C}$~\cite[Thm. 5.6.26]{kad}.  If
$\mathtt{C}$ is a Borel subset of $\spectrum{R}$, then $E(\mathtt{C})$
is defined to be $\chi _{\mathtt{C}}(R)$, where $\chi _{\mathtt{C}}$
is the characteristic function of $\mathtt{C}$.  If $\mathtt{C}$ is
any Borel subset of $\mathbb{C}$, then $E(\mathtt{C})$ is defined to
be $E(\mathtt{C}\cap \spectrum{R})$.  Note that for any
$\mathtt{C}\subseteq \mathbb{C}$, $\phi (E(\mathtt{C}))$ is a
characteristic function (since $E(\mathtt{C})$ is a projection) and is
actually \emph{continuous} on $S$ (since $\phi (E(\mathtt{C}))\in
\mathcal{C}(S)$).

In what follows, we specialize to the case where $R$ is
\emph{self-adjoint}, so that $\spectrum{R}\subseteq \mathbb{R}$, and
$\phi (R)$ is a self-adjoint function on $\states{S}$.  We consider
$\spectrum{R}$ with the order relation inherited from $\mathbb{R}$ and
with the relative topology inherited from $\mathbb{R}$.  Recall that a
\emph{convex} subset of $\spectrum{R}$ is any subset $\reals{C}$ with
the following property: If $a,b\in \reals{C}$ and there is a $c\in
\spectrum{R}$ such that $a<c<b$, then $c\in \reals{C}$.  Note that the
relative basis of $\spectrum{R}$ consists of convex sets with compact
closure.  If $\reals{C}\subseteq \spectrum{R}$, we let $\clo{C}$
denote the closure of $\reals{C}$ with respect to the relative
topology.

\begin{lemma} Let $\omega$ be a pure state of $\vn{R}$, and let
  $\reals{C}$ be a convex subset of $\spectrum{R}$ with compact
  closure.  \begin{enumerate}
  \item If $\omega (E(\reals{C}))=1$, then $\phi (R)$ is defined at
    $\omega$ and $\phi(R) (\omega )\in \clo{C}$.
  \item If $\reals{C}$ is open and $\phi (R)(\omega )\in \reals{C}$,
    then $\omega (E(\reals{C}))=1$.  \end{enumerate} \label{yes}
\end{lemma}

\begin{proof} (i) Suppose that $\omega (E(\reals{C}))=1$, and consider $\chi _{\reals{C}}\in
  \mathcal{B}_{u}(\spectrum{R})$, the characteristic function of
  $\reals{C}$.  Define $\widetilde{\chi _{\reals{C}}}\in
  \mathcal{B}_{u}(\states{S})$ as in~(\ref{func_calc}), so
  $\widetilde{\chi _{\reals{C}}}(\omega )=1$ if $\phi (R)(\omega )\in
  \reals{C}$, $=0$ otherwise.  Let $h$ be the unique function in
  $\mathcal{C}(\states{S} )$ that agrees with $\widetilde{\chi
    _{\reals{C}}}$ on the complement of a meager set.  Thus,
  $E(\reals{C})\equiv \phi ^{-1}(h)$ and $h(\omega )=\omega
  (E(\reals{C}))=1$.
  
  Suppose, for reductio ad absurdum, that $\phi (R)$ is not defined at
  $\omega$, so that $\widetilde{\chi _{\reals{C}}}(\omega )=0$.  Since
  $\phi (R)$ is self-adjoint, $\lim_{\tau \rightarrow \omega}\abs{\phi
    (R)(\tau )}\rightarrow \infty$.  Since $\reals{C}$ is bounded,
  there is an open neighborhood $\states{U}$ of $\omega$ such that
  $\phi (R)(\tau )\not\in \reals{C}$, for all $\tau\in \states{U}$.
  Thus, $\widetilde{\chi _{\reals{C}}}(\states{U})=\{ 0\}$.  However,
  since $h$ is a continuous map from $\states{S} $ into $\{ 0,1\}$ and
  $h(\omega )=1$, there is an open neighborhood $\states{V}$ of
  $\omega$ such that $h(\states{V})=\{ 1\}$.  But then
  $\widetilde{\chi _{\reals{C}}}$ and $h$ disagree on the non-empty
  open set $\states{U}\cap \states{V}$, which is impossible.
  Therefore, $\phi (R)$ is defined at $\omega$.
  
  Again, suppose for reductio that $\phi (R)$ is defined at $\omega$
  but that $\phi (R)(\omega )\not\in \reals{C}^{-}$.  Since $\phi (R)$
  is defined at $\omega$, it is continuous at $\omega$. Hence $\omega
  \not\in [\phi (R)^{-1}(\reals{C})]^{-}$ and there is an open
  neighborhood $\states{U}$ of $\omega$ such that $\states{U}\cap
  [\phi (R)^{-1}(\reals{C})]=\emptyset$.  Thus, $\phi
  (R)(\states{U})\cap \reals{C}=\emptyset$ and $\widetilde{\chi
    _{\reals{C}}}(\states{U})=\{ 0\}$.  Since $h$ is continuous, there
  is an open neighborhood $\states{V}$ of $\omega$ such that
  $h(\states{V})=\{ 1\}$.  But then $\widetilde{\chi _{\reals{C}}}$
  and $h$ disagree on the non-empty open set $\states{U}\cap
  \states{V}$, which again is impossible.  Therefore $\phi (R)(\omega
  )\in \clo{C}$.
  
  (ii) Suppose that $\reals{C}$ is open and $\phi(R)(\omega )\in
  \reals{C}$.  Let $h\equiv \phi (E(\reals{C}))$.  We must show that
  $h(\omega )=1$.  Recall from (i) that $h$ agrees on the complement
  of a meager set with $\widetilde{\chi _{\reals{C}}}\in
  \mathcal{B}_{u}(\states{S} )$.  By assumption, then,
  $\widetilde{\chi _{\reals{C}}}(\omega )=1$.  Suppose, for reductio,
  that $h(\omega )=0$.  Since $h$ is continuous, there is an open
  neighborhood $\states{U}$ of $\omega$ such that $h(\states{U})=\{
  0\}$.  Since $\phi (R)$ is continuous on $\states{S}
  \backslash\states{Z}$, $\states{V}\equiv \phi (R)^{-1}(\reals{C})$
  is open in $\states{S} \backslash\states{Z}$ (and thus open in
  $\states{S}$, since $\states{S} \backslash\states{Z}$ is open in
  $\states{S}$).  Then, $\widetilde{\chi _{\reals{C}}}(\states{V})=\{
  1\}$ and $\states{U} \cap \states{V}$ (which contains $\omega$) is a
  non-empty open set on which $h$ and $\widetilde{\chi _{\reals{C}}}$
  disagree---a contradiction.  Therefore, $h(\omega )=\omega
  (E(\reals{C}))=1$.  \end{proof}

The next proposition confirms what might otherwise expect: that $R$
may be assigned a dispersion-free value $\ell \in \spectrum{R}$
exactly when all propositions of the form `the value of $R$ lies in
$\reals{K}$', for all compact convex $\reals{K}\subset \spectrum{R}$
that contain $\ell$, are true.  (Clearly if this were not so---in
particular, if \emph{no} proposition of that form were true---then it
would make no physical sense to assign $R$ any value whatsoever.)

\begin{prop} Let $\omega$ be a pure (dispersion-free) state of
  $\vn{R}$.  Then $\phi (R)$ is defined at $\omega$ if and only if
  there is a compact convex set $\reals{K}\subset \spectrum{R}$ such
  that $\omega (E(\reals{K}))=1$.  If these conditions hold, then
  \begin{equation*} \{ \phi (R)(\omega )\} =\bigcap \Bigl\{ \reals{K} :\reals{K}\;
    \text{is a compact convex set in $\spectrum{R}$ and $\omega
      (E(\reals{K}))=1$} \Bigr\} \,. \end{equation*} \end{prop}

\begin{proof} ``$\Leftarrow$''  Immediate from Lemma~\ref{yes}.(i).
  
  ``$\Rightarrow$'' Suppose that $\phi (R)$ is defined at $\omega \in
  \states{S}$.  Then, since there is an open convex neighborhood
  $\reals{C}$ of $\phi (R)(\omega )$ such that $\clo{C}$ is compact
  (and convex), $\omega (E(\reals{C}))=1$~(by Lemma~\ref{yes}.(ii)).
  Moreover, since a projection-valued measure is monotone and states
  are order-preserving, \begin{equation} 1\;=\;\norm{E(\clo{C})}\;\geq
    \;\omega (E(\clo{C}))\;\geq \;\omega (E(\reals{C}))\;=\;1,
  \end{equation} which entails $\omega (E(\clo{C}))=1$.
  
  Suppose now that the the above equivalent conditions hold for $\phi
  (R)$ and $\omega$, and let $\reals{Y}$ denote the intersection in
  the statement of this proposition.  By assumption, there is at least
  one compact convex set $\reals{K}$ such that $\omega
  (E(\reals{K}))=1$, so that $\reals{Y}$ is nonempty. Let $\reals{L}$
  be any other such set where $\omega (E(\reals{L}))=1$.  Then, by
  Lemma~\ref{yes}.(i), $\phi (R)(\omega )\in \clo{L}=\reals{L}$.
  Therefore, $\phi (R)(\omega )\in \reals{Y}$.
  
  Finally, to see that $\phi (R)(\omega )$ is the unique point in
  $\reals{Y}$, suppose that $\lambda \in \reals{Y}$ yet $\lambda \neq
  \phi (R)(\omega )$.  Since $\spectrum{R}$ is Hausdorff, there is an
  open convex neighborhood $\reals{C}$ in $\spectrum{R}$ such that
  $\phi (R)(\omega )\in \reals{C}$ but $\lambda \not\in \clo{C}$.  (We
  choose $\reals{C}$ such that its closure is compact.)  Since $\phi
  (R)(\omega )\in \reals{C}$ and $\reals{C}$ is open, $\omega
  (E(\clo{C}))=1$ (by Lemma~\ref{yes}.(ii)).  Therefore, $\lambda
  \not\in \reals{Y}$, contradicting our assumption.  It follows that
  $\phi (R)(\omega )$ is the unique element of $\reals{Y}$.
\end{proof}

Given a von Neumann algebra $\alg{B}$, beable for a state $\rho$, it
is natural to ask when an unbounded self-adjoint operator $R$ (or a
family of such observables) affiliated with $\alg{B}$ can be taken to
have beable status for $\rho$ \emph{together with the observables in}
$\alg{B}$.  This would require that $\rho$ be a mixture of
dispersion-free states on $\alg{B}$ each of which restricts to a pure
state on $\vn{R}\subseteq\alg{B}$ that permits a value for $R$ to be
defined in accordance with the above proposition.  As we show in
Theorem~\ref{thm_beableaff} below, a sufficient condition for this is
that $\rho$ determine a finite expectation value for $R$.

While pure states of $\vn{R}$ correspond to points of $\states{S}$,
the general state $\rho$ (pure or mixed) of $\vn{R}$ corresponds
uniquely, via the Riesz Representation Theorem, to a probability
measure $\mu _{\rho}$ on $\states{S}$.  (A pure state corresponds to a
measure concentrated at a single point.)  That is, \begin{xalignat}{2}
  \rho (A)&=\int_{\states{S}}\phi (A)(s)d\mu _{\rho}(s) &(A\in
  \vn{R}).
\end{xalignat}
Since $\phi (R)$ is an unbounded function on $\states{S}$, its
integral with respect to $\mu _{\rho}$ may or may not converge to a
finite value.  In order to capture the idea that some states may be
used consistently to assign finite (not necessarily dispersion-free)
expectation values to unbounded operators, we introduce the following
notion of a well-defined state:
\begin{defn} Suppose that $\rho$ is a state of $\vn{R}$ and
  that $\mu _{\rho}$ is the measure on $\states{S}$ corresponding to
  $\rho$.  If $\int_{\states{S} }\phi (R)d\mu _{\rho}<\infty$, we say
  that $\rho$ is a \emph{well-defined} state for $R$.  \label{defined}
\end{defn} 
\noindent As should be the case, this definition entails that if $\rho$ is
a pure state, then $\rho$ is well-defined for $R$ if and only if $\phi
(R)$ is defined at $\rho$.  Moreover, by~\cite[Thm. 5.6.26]{kad}, a
vector state $\omega _{x}$ is well-defined for $R$ if and only if $x$
is in the domain of $R$.  Of course, this definition may easily be
extended to any von Neumann algebra $\alg{V}$, such that $R$ is
affiliated with $\alg{V}$.  If $\rho$ is a state of $\alg{V}$, we say
that $\rho$ is well-defined for $R$ just in case $\rho \,|_{\vn{R}}$
is well-defined for $R$.

\begin{remark} Of course it is possible for $\rho$ to be well-defined
  for $R$, but not for polynomials in $R$.  For example, let $R=Q$ be
  the the multiplication by $x$ operator on $L_{2}(\mathbb{R})$.
  Then, one can easily construct unit vectors in
  $\hil{D}(Q)-\hil{D}(Q^{2})$ whose corresponding states will be
  well-defined for $Q$ but not for $Q^{2}$. \end{remark}
\noindent Pure well-defined states, however, are extremely well-behaved:
\begin{prop}[Cont-FUNC] Let $R$ be a (possibly unbounded) self-adjoint
  operator on $\hil{H}$ and suppose that $\omega$ is a pure state of
  $\bh$ that is well-defined for $R$.  Then, $\omega (f(R))=f(\omega
  (R))$, for any $f\in \mathcal{C}(\spectrum{R})$.  \end{prop}

\begin{proof} Note that
  $\widetilde{f}\,|_{\states{S}\backslash\states{Z}}$ is continuous,
  being the composition of two continuous functions, $f$ and $\phi
  (R)\,|_{\states{S}\backslash\states{Z}}$.  Moreover, since the
  normal function $h$ agrees with $\widetilde{f}$ on the complement of
  a meager set, $h$ must agree with $\widetilde{f}$ throughout
  $\states{S}\backslash\states{Z}$.  Thus, $\omega
  (f(R))=h(\omega)=\widetilde{f}(\omega )=f(\omega (R))$.  \end{proof}
 
\begin{lemma} Let $\rho$ be any state (pure or mixed) of
  $\vn{R}$, and let $F_{n}\equiv E((-n,n])$, where $E$ is the
  projection-valued measure associated with $R$.  If $\rho$ is
  well-defined for $R$, then $\lim_{n\rightarrow \infty}\rho
  (F_{n})=1$.
\label{rho_converge} \end{lemma}

\begin{proof} Let $\widetilde{g}_{n}$ be defined as
  $\widetilde{g}_{n}(\omega )=1$ if $\phi (R)(\omega )\in (-n,n]$ and
  $\widetilde{g}_{n}(\omega )=0$ otherwise.  Then, $F_{n}\equiv \phi
  ^{-1}(h_{n})$ where $h_{n}$ is the unique function in
  $\mathcal{C}(\states{S})$ which agrees with $\widetilde{g}_{n}$ on
  the complement of a meager set.  Clearly then, $\{ h_{n}\}$
  converges pointwise to $\chi _{(\states{S}\backslash\states{Z})}$,
  the characteristic function of $\states{S}\backslash\states{Z}$, as
  $n\rightarrow \infty$. Thus, $\rho
  (F_{n})=\int_{\states{S}}h_{n}d\mu _{\rho}$, and
  \begin{equation} \lim_{n\rightarrow \infty}\rho
    (F_{n})\:=\:\lim_{n\rightarrow \infty}\int_{\states{S}}h_{n}d\mu
    _{\rho}\:=\:\int_{\states{S}}\chi _{(\states{S}\backslash\states{Z})}d\mu _{\rho}
    \:=\:\mu _{\rho}(\states{S}\backslash\states{Z}), \end{equation} where the second
  equality follows from the Monotone Convergence Theorem~\cite[Thm.
  1.26]{rud}.  Hence, $\mu _{\rho}(\states{Z})=1-\lim_{n\rightarrow \infty}\rho
  (F_{n})$.
  
  Since $\phi (R)$ is a self-adjoint function on $X$, we may decompose
  $\states{Z}$ as $\states{Z}_{+}\cup \states{Z}_{-}$ where
  $\states{Z}_{+}$ is the set of points $\omega$ of $\states{S}$ such
  that $\lim_{\tau \rightarrow \omega}\phi (R)(\tau )=+\infty$ and
  $\states{Z}_{-}$ is the set of points $\omega$ of $\states{S}$ such
  that $\lim_{\tau \rightarrow \omega }\phi (R)(\tau
  )=-\infty$~\cite[p. 344]{kad}.  Now, let $f_{+}\equiv \max \{ \phi
  (R),0\}$ be the positive part of $\phi (R)$ and let $f_{-}\equiv
  -\min \{ \phi (R),0 \}$ the negative part.  Then,
\begin{align}
  \int_{\states{S}}f_{+}d\mu _{\rho}
  &=\int_{(\states{S}\backslash\states{Z}_{+})}f_{+}d\mu _{\rho}
  +\int_{\states{Z}_{+}}f_{+}d\mu _{\rho} \\
  &=\int_{(\states{S}\backslash\states{Z}_{+})}f_{+}d\mu _{\rho}+(\mu
  _{\rho} (\states{Z}_{+})\cdot +\infty),
\label{positive} \\
\intertext{and similarly,} \int_{\states{S}}f_{-}d\mu _{\rho}
&=\int_{(\states{S}\backslash\states{Z}_{-})}f_{-}d\mu _{\rho}+(\mu
_{\rho}(\states{Z}_{-})\cdot
-\infty). \label{negative} \\
\intertext{By definition, $\int_{\states{S}}\phi(R)d\mu _{\rho}$ is
  defined only if either~(\ref{positive}) or~(\ref{negative}) is
  finite, and then,} \int_{\states{S}}\phi (R)d\mu _{\rho}&\equiv
\int_{\states{S}}f_{+}d\mu _{\rho}-\int_{\states{S}}f_{-}d\mu _{\rho}.
\end{align}
Thus, $\mu _{\rho}(\states{Z})=\mu _{\rho}(\states{Z}_{+})+\mu
_{\rho}(\states{Z}_{-})$, and if $\mu _{\rho}(\states{Z})>0$ then
either~(\ref{positive}) or~(\ref{negative}) is infinite and either
$\int_{\states{S}}\phi (R)d\mu _{\rho}$ is undefined or $=\pm \infty$.
Therefore, $\int_{\states{S}}\phi (R)d\mu _{\rho}$ has a finite value
only if $\lim_{n\rightarrow \infty}\rho (F_{n})=1$.  \end{proof}

\begin{thm} Suppose $\alg{B}$ is a von Neumann algebra and
  $\alg{B}$ is beable for $\rho$.  Suppose that $\{ R_{j}\}$ is a
  countable family of self-adjoint operators affiliated with $\alg{B}$
  such that, for all $j\in \mathbb{N}$, $\rho$ is a well-defined state
  for $R_{j}$.  Then, there is a probability measure $\mu$ on the set
  of dispersion-free states $\states{S}$ of $\alg{B}$ such that
  \begin{xalignat}{2} \rho (A) &=\int_{\states{S} }\omega _{s}(A)d\mu (s) &(A\in
    \alg{B}), \end{xalignat} and for every $\omega _{s}\in \states{S}$
  and $j\in \mathbb{N}$, $\omega _{s}$ is well-defined for $R_{j}$.
\label{thm_beableaff} \end{thm}

\begin{proof} Since $\alg{B}$ is beable for $\rho$, we have a
  probability measure $\mu$ on the set of dispersion-free states
  $\states{T}$ of $\alg{B}$ such that \begin{xalignat}{2} \rho (A)
    &=\int_{\states{T}}\omega _{t}(A)d\mu (t) &(A\in \alg{B}).
  \end{xalignat} We will show that $\mu (\states{S} )=1$, where $\states{S}$
  is the subset of $\states{T}$ consisting of those states that are
  well-defined for each $R_{j}$.
  
  Fix $j\in \mathbb{N}$.  Let $\vn{R_{j}}$ be the von Neumann algebra
  generated by $R_{j}$, let $E^{j}$ be the projection-valued measure
  on $\mathbb{R}$ induced by $R_{j}$, and let $F^{j}_{n}\equiv
  E^{j}((-n,n])$.  Let
\begin{equation}
\states{Z} _{j}\equiv \{ t\in \states{T}: \omega _{t}(F^{j}_{n})=0, \,\text{for all $n\in
\mathbb{N}$} \}.
\end{equation} It is not difficult to verify that $\states{Z} _{j}$ is a measurable
subset of $\states{T}$.  Suppose, for reductio, that $\mu (\states{Z}
_{j})=\delta >0$ so that $\mu (\states{T}\backslash \states{Z}
_{j})=1-\delta $.  Choose any $m\in \mathbb{N}$.  Then,
\begin{align} \rho (F^{j}_{m}) &= \int_{\states{Z} _{j}} \omega_{t}(F^{j}_{m})d\mu (t)
  +\int_{(\states{T}\backslash \states{Z} _{j})} \omega _{t}(F^{j}_{m})d\mu (t) \\
  &=0+\int_{(\states{T}\backslash \states{Z} _{j})} \omega _{t}(F^{j}_{m})d\mu (t) \label{bstates} \\
  &\leq \mu (\states{T}\backslash \states{Z} _{j})
  \label{two_values}=1-\delta.
\end{align}
(\ref{bstates})~follows by the definition of $\states{Z} _{j}$, and
the inequality in~(\ref{two_values}) follows since $\omega
_{t}(F^{j}_{m})\leq \norm{F^{j}_{m}}=1$ for all $t\in \states{T}$.
Since $m$ was arbitrary $\lim_{n\rightarrow \infty}\rho
(F^{j}_{n})\leq 1-\delta$.  By Proposition~\ref{rho_converge}, $\rho
|_{\alg{A}_{j}}$ does not correspond to a convergent measure,
contradicting our assumption that $\rho$ is well-defined for $R_{j}$.
Thus, $\mu (\states{Z} _{j})=0$.  Since $\mu$ is countably additive,
$\mu (\cup_{j=1}^{\infty}\states{Z} _{j})=0$.

Let $\states{S} =\states{T}\backslash (\cup_{j=1}^{\infty} \states{Z}
_{j})$.  Then, for $A\in \alg{B}$, \begin{align} \rho (A) &=
  \int_{\states{T}}\omega _{t}(A)d\mu (t)\:=\:\int_{\states{S} }\omega
  _{t}(A)d\mu (t)
  +\int_{\cup \states{Z} _{j}}\omega _{t}(A)d\mu (t) \\
  &=\int_{\states{S} }\omega _{t}(A)d\mu (t)\:=\:\int_{\states{S}
    }\omega _{s}(A)d\mu (s),
\end{align} where the penultimate equality follows since $\mu
(\cup_{j=1}^{\infty} \states{Z} _{j})=0$.  Finally, suppose that
$\omega \in \states{S}$; that is, for each $j$ there is an $m$ such
that $\omega (F^{j}_{m})\equiv \omega (E^{j}((-m,m]))>0$.  But $\omega
(F^{j}_{m})\in \{ 0,1\}$ since $\omega$ is dispersion-free on
$\alg{A}_{j}$ and $F^{j}_{m}$ is a projection.  Thus, for each $j$
there is an $m$ such that $\omega (F^{j}_{m})=1$, and by
Lemma~\ref{yes}.(i), $\omega$ is well-defined for each $R_{j}$.
Therefore, $\rho$ is a mixture of dispersion-free states of $\alg{B}$,
all of which are well-defined for each $R_{j}$.  \end{proof}

\begin{defn} Let $\{ R_{\lambda}:\lambda\in\Lambda\}$ be a family of 
  (possibly unbounded) self-adjoint operators acting on a Hilbert
  space $\hil{H}$, and $\rho$ a state of $\bh$.  We say that the
  observables $\{ R_{\lambda}:\lambda\in\Lambda\}$ have \emph{joint
    beable status} for $\rho$ if there is a subalgebra
  $\alg{B}\subseteq\bh$, to which each $R_{\lambda}$ is affiliated,
  such that $\rho$ is a mixture of dispersion-free states on $\alg{B}$
  $\mu_{\rho}$-measure-one of which are well-defined for each
  $R_{\lambda}$.  Thus, a family of observables has joint beable
  status in a state just in case it is possible to think of the
  observables as possessing simultaneously determinate values without
  contradicting the state's expectation values.  In particular, when
  $\rho$ is well-defined on all the observables $\{
  R_{\lambda}:\lambda\in\Lambda\}$, their joint beable status for
  $\rho$ (sufficient conditions for which are identified in Theorem
  \ref{thm_beableaff} above) guarantees that the expectation values
  $\rho$ assigns to each $R_{\lambda}$ can be interpreted as arising
  due to ignorance about the precise values jointly possessed by the
  observables in $\{ R_{\lambda}:\lambda\in\Lambda\}$.
\end{defn}

As should be the case, the bounded observables in any subalgebra
$\alg{B}\subseteq\bh$ beable for $\rho$ have joint beable status for
$\rho$.  And, of course, any single bounded observable $R$---being
affiliated with the abelian von Neumann algebra it generates---has
beable status in any state.  However, when $R$ is unbounded, this need
not be true, as the next results show.

\begin{prop}  Let $A,B$ be canonically conjugate self-adjoint unbounded operators on 
  some Hilbert space $\hil{H}$, that is, they satisfy $[A,B]=\pm iI$
  with $\spectrum{A}=\spectrum{B}=\mathbb{R}$.  Let $\rho$ be a state
  of $\bh$ such that $\rho \,|_{\vn{A}}$ is pure, and $\rho$ is
  well-defined for $A$.  Then $\rho (E(\reals{C}))=0$ for any compact
  interval $\reals{C}$ in $\mathbb{R}$, where $E$ is the
  projection-valued measure for $B$.  \label{prop_cool} \end{prop}

\begin{proof}  We show first that $\rho (\cos tB)=0$ for all $t\in
  \mathbb{R}\backslash \{ 0\}$.  For this, let $U_{s}\equiv e^{isA}$
  and let $W_{t}\equiv e^{itB}$.  Then, invoking the Weyl form of
  $[A,B]=\pm iI$ (taking either sign), we have $U_{s}W_{t}=e^{\pm
    ist}W_{t}U_{s}$ for all $s,t\in \mathbb{R}$.  Thus, $\rho
  (U_{s}W_{t})=e^{\pm ist}\rho (W_{t}U_{s})$.  Moreover, since
  $U_{s}\in \vn{A}$, $\rho$ is dispersion-free on $U_{s}$ and $\rho
  (U_{s})\rho (W_{t})=e^{\pm ist}\rho (W_{t})\rho (U_{s})$.  Again,
  since $\rho$ is dispersion-free on $\vn{A}$, $\rho (U_{s})\neq 0$
  for all $s\in \mathbb{R}$, and $\rho (W_{t})=e^{\pm ist}\rho
  (W_{t})$ for all $s,t\in \mathbb{R}$.  Let $t=t_{0}\neq 0$.  Then we
  may choose $s$ such that $e^{\pm ist_{0}}\neq 1$, and hence $\rho
  (W_{t_{0}})=0$.  But $t_{0}$ was an arbitrary non-zero number; thus,
  $\rho (W_{t})=0$ for all $t\neq 0$.  Moreover, $\rho (W_{t})=\rho
  (\cos tB)+i\rho (\sin tB)$, from which it follows that $\rho (\cos
  tB)=0$ for all $t\neq 0$.
  
  Recall that \begin{equation} \cos
    ^{2n}\theta=\frac{1}{2^{2n}}\binom{2n}{n}+
    \frac{1}{2^{2n-1}}\sum_{m=0}^{n-1}\binom{2n}{m}\cos 2(n-m)\theta.
    \label{cosine}
    \end{equation}
    Let $F_{t}^{n}\equiv \cos ^{2n}tB$.  From~(\ref{cosine}) we may
    deduce the operator identity:
\begin{equation}  \cos ^{2n}tB=\frac{1}{2^{2n}}\binom{2n}{n}+
    \frac{1}{2^{2n-1}}\sum_{m=0}^{n-1}\binom{2n}{m}\cos 
    2(n-m)tB. \end{equation}
Thus, from the linearity of $\rho$, in combination with the result of the
previous paragraph, we may conclude that $\rho
(F_{t}^{n})=2^{-2n}\binom{2n}{n}\equiv k(n)$ whenever $t\neq 0$.  And, 
using Stirling's approximation for the factorial, $k(n)\approx (\pi 
n)^{-1/2}$ for large $n$, whence $\lim_{n\rightarrow\infty}k(n)=0$.

Now let $\reals{C}$ be a compact interval in $\mathbb{R}$.  Then,
$\rho (F_{t}^{n}E(\reals {C}))\leq \rho (F_{t}^{n})=k(n)$, for all
$n\in \mathbb{N}$ and all $t\neq 0$.  Consider the extended function
representation $\mathcal{N}(\states{S})$ of the abelian von Neumann
algebra $\vn {B}$.  Let $\states{T}\equiv \{\omega \in
\states{S}:\omega (E(\reals{C}))=1\}$.  ($\states{T}$ is clopen since
it is the support of the continuous idempotent function $\phi
(E(\reals{C}))$.)  Fix $n\in \mathbb{N}$ and let $f_{t}\equiv \phi
(F_{t}^{n}E(\reals{C}))$ for each $t\in \mathbb{R}\backslash \{ 0\}$.
We claim that $f_{t}$ converges pointwise to $\chi _{\states{T}}$ as
$t\!\rightarrow\!0$.  Note first that $\phi (B)$ is defined at all
points of $\states{T}$ and $\phi (B)(\states{T})\subseteq \reals{C}$
by Lemma~\ref{yes}.(i).  Now, for any $\epsilon >0$, we may choose $t$
small enough that $1-\cos ^{2n}tx<\epsilon$ for all $x\in \reals{C}$
(since $\reals{C}$ is compact and $n$ is fixed).  Thus, for any
$\omega \in \states{T}$ (and using Cont-FUNC in the fourth step),
\begin{align} 
  \chi _{\states{T}}(\omega )-f_{t}(\omega ) & =
  1-\phi(F_{t}^{n}E(\reals{C}))(\omega) =
  1-\omega(F_{t}^{n}E(\reals{C})) \\ & = 1-\omega(F_{t}^{n}) =
  1-\cos^{2n}(t\omega(B)) = 1-\cos ^{2n}(t\phi (B)(\omega
  ))\:<\:\epsilon ,\end{align} which is what we needed to show.

Since $f_{t}$ converges pointwise to $\chi _{\states{T}}$ we may apply
the Dominated Convergence Theorem~\cite[Thm. 1.34]{rud} to conclude
that \begin{equation} \lim _{t\rightarrow 0}\int
  _{\states{S}}f_{t}d\mu _{\rho}\:=\:\int _{\states{S}}\chi
  _{\states{T}}d\mu _{\rho}\:=\:\mu _{\rho}(\states{T})\:=\:\rho
  (E(\reals{C})) .\end{equation} However, since $k(n)\geq \rho
(F_{t}^{n}E(\reals {C}))=\int f_{t}d\mu _{\rho}$, for all $t\neq 0$,
it follows that $k(n)\geq \rho (E(\reals{C}))$.  Since this is true
for all $n\in \mathbb{N}\,$, and $\lim_{n\rightarrow\infty}k(n)=0$, it
follows that $\rho (E(\reals{C}))=0$.  \end{proof}

\begin{cor} Let $A,B$ be as above.  Then $\mu _{\rho}(\states{Z})=1$, where $\states{Z}$
  is the set of states at which $B$ is not defined.  In particular,
  when $\rho$ is a state of $\bh$ such that $\rho \,|_{\vn{A}}$ is
  pure and $\rho$ is well-defined for $A$, then $B$ does \emph{not}
  have beable status for $\rho$.  \label{cor_cool}\end{cor}

\begin{proof} Let $E_{n}\equiv E([-n,n])$.  Let $\states{S_{n}}\equiv
  \{\omega \in \states{S}:\omega (E_{n})=1\}$.  Then, from the
  preceding Proposition, $\rho (E_{n})=0$, and thus $\mu
  _{\rho}(\states{S}_{n})=0$, for all $n\in \mathbb{N}$.  However,
  $\cup _{n=1}^{\infty}\states{S}_{n}=\states{S}\backslash
  \states{Z}$, and it follows from the countable additivity of $\mu
  _{\rho}$ that $\mu _{\rho}(\states{S}\backslash \states{Z})=0$.
\end{proof}

\begin{example}[Heisenberg-Bohr Indeterminacy Principle] Let $D$ and $Q$ 
  be the momentum and position operators for a particle in
  one-dimension with state space $L_{2}(\mathbb{R})$.  It is a
  well-known consequence of $[Q,D]=i\hbar I$ that the product of the
  dispersions of $Q$ and $D$, for all wavefunctions
  $\psi\in\mathcal{D}(QD)\cap \mathcal{D}(DQ)$, is bounded below by
  $\hbar$.  The standard Copenhagen interpretation of this uncertainty
  principle is not simply that a precision momentum measurement
  necessarily and uncontrollably disturbs the value of position, and
  vice-versa, but that $D$ and $Q$ can never in reality be thought of
  as simultaneously determinate.  The warrant for this stronger
  `indeterminacy principle' is not obvious, since there appears to be
  nothing preventing the view that the dispersion required in (say) a
  particle's momentum when its position is measured simply reflects
  our loss of knowledge about that momentum---not any breakdown in the
  applicability of the momentum concept itself.  However, the
  foregoing results allow us to exhibit the indeterminacy principle as
  a direct mathematical consequence of $[Q,D]=i\hbar I$ (and without
  taking any a priori stand on precisely which (if any) of the many
  subalgebras with beable status for a given state should be taken to
  represent observables that \emph{actually} possess determinate
  values).  As we have seen, a necessary (and sufficient) condition
  for thinking of $Q$ and $D$ as having simultaneously determinate
  values in a state $\rho$ is that they have joint beable status for
  $\rho$.  This, in turn, requires that $\rho$ be a mixture of states
  (on some subalgebra of $\alg{L}(L_{2})$) each of which is pure on
  both $\vn{Q}$ and $\vn{D}$ and well-defined on both $Q$ and $D$.
  Yet, as Proposition \ref{prop_cool} and its Corollary make clear,
  satisfaction of these requirements for $Q$ precludes their
  satisfaction for $D$, and vice-versa.  It follows that there is
  \emph{no} state $\rho$ for which $Q$ and $D$ have joint beable
  status, and the indeterminacy principle is proved.\end{example}

\section{Beable Subalgebras Determined by a Family of Privileged Observables}
It is evident from Theorem~\ref{thm_def} that any subspace
$\hil{T}\subseteq\hil{H}$ containing $\hil{K}$, together with any
maximal abelian subalgebra of $\alg{L}(\hil{T})$, determines a maximal
beable subalgebra $\alg{B}\subseteq\alg{L}(\hil{H})$ for $K$.  In the
present section we take steps to eliminate this arbitrariness.  Let
$\alg{A}$ be a $C^{*}$-algebra and let $\alg{R}$ be a mutually
commuting family of ``privileged'' observables drawn from $\alg{A}$.
We may then inquire into the structure of all beable algebras for a
given state that contain the commuting family $\alg{R}$.

The reasons why one might want to demand a priori that certain
preferred observables $\alg{R}$ be included in the subalgebra with
beable status will become apparent when we apply our results to the
orthodox Copenhagen interpretation of quantum theory below.  We shall
also be requiring that a beable subalgebra $\alg{B}$ for $\rho$
containing some set of observables $\alg{R}$ be (at least implicitly)
definable in terms of $\alg{R}$, $\rho$, and the algebraic operations
available within $\alg{A}$.  This idea is captured by requiring that
$\alg{B}$ be invariant under spatial automorphisms of $\alg{A}$ that
fix both $\alg{R}$ and the state $\rho$.  (We say that the spatial
automorphism $\Phi$ induced by unitary $U$ \emph{fixes} $\rho$ just in
case $\rho_{U}=\rho$, where $\rho_{U}$ is defined by
$\rho_{U}(A)=\rho(U^{*}AU)$ for all $A$ in $\alg{A}$.)
\begin{defn} Let $\alg{A}$ be a $C^{*}$-algebra, let $\alg{R}$ be
  any mutually commuting family of observables in $\alg{A}$, and let
  $\rho$ be a state of $\alg{A}$.  Then, for any subalgebra $\alg{B}$
  of $\alg{A}$, we say that $\alg{B}$ is \emph{$\alg{R}$-beable} for
  $\rho$ just in case:
\begin{description}
\item[(Beable)] $\alg{B}$ is beable for $\rho$.
\item[($\alg{R}$-Priv)] $\alg{R}\subseteq\alg{B}$.
\item[(Def)] For any unitary $U\in \alg{A}$, if $U\in \alg{R}'$ and
  $\rho _{U}=\rho$, then $U\alg{B}U^{*}=\alg{B}$.
\end{description} We say that $\alg{B}$ is \emph{maximal $\alg{R}$-beable}
for $\rho$ if and only if $\alg{B}$ is maximal with respect to the
properties (Beable), ($\alg{R}$-Priv), and (Def) (noting that, by
Zorn's lemma, maximal $\alg{R}$-beable algebras exist for any state).
\label{defn:rbeable}
\end{defn}

\subsection{$\alg{R}$-beable algebras for normal states.}
We now specialize to the case where $\alg{A}=\bh$, and where $\rho$ is
a normal state of $\bh$.  In this case, we may replace $\rho
_{U}=\rho$ in (Def) by $UKU^{*}=K$, where $K$ is the trace-$1$
operator that defines the state $\rho$.  We shall soon see that the
above requirements, for certain $\alg{R}$ and $K$, suffice to
determine a \emph{unique} maximal $\alg{R}$-beable algebra for
$K$~(cf.~Corollary~\ref{unique} below).

\begin{lemma} Suppose that $\alg{B}$ is a $C^{*}$-algebra acting on some Hilbert
  space $\hil{H}$, and that $\rho$ is a normal state of $\alg{B}$.  If
  $\alg{B}$ is $\alg{R}$-beable for $\rho$, then $\alg{B}^{-}$ is also
  $\alg{R}$-beable for $\rho$.  \label{juji}
\end{lemma}

\begin{proof}
  ($\alg{R}$-Priv) $\alg{R}\subseteq \alg{B} \subseteq \alg{B}^{-}$.
  (Beable) See Corollary~\ref{cor_woclosed}.(ii).  (Def) Suppose that
  $U$ is a unitary element of $\bh$ such that $U\in \alg{R}'$ and
  $UKU^{*}=K$.  Then, since $\alg{B}$ satisfies (Def), $U\alg{B}
  U^{*}=\alg{B}$.  Since the spatial automorphism $\Phi$ of $\bh$
  induced by $U$ is a WOT-homeomorphism from $\alg{L}(\hil{H})$ to
  $\alg{L}(\hil{H})$, it follows that
  $\Phi(\alg{B^{-}})=\Phi(\alg{B})^{-}=\alg{B}^{-}$.
\end{proof}

In order to prove the main result of this section, we will need to
make use of the following lemma:

\begin{lemma} Let $Q\in \alg{L}(\hil{H} )$ be a projection, and let
  $\alg{V}$ be a von Neumann algebra acting on $\hil{H}$.  Suppose
  that for every unitary operator $U\in \alg{V}\,'$, $[UQU^{*},Q]=0$.
  Then, $Q\in \alg{V}$. \label{lemma_stone}
\end{lemma}

\begin{remark} Recall that every element of a $C^{*}$-algebra (such 
  as $\alg{V}\,'$) is expressible as a linear combination of (four)
  unitary elements in that algebra~\cite[Theorem 4.1.7]{kad}.  Thus,
  we may reformulate Lemma~\ref{lemma_stone} equivalently as: If
  $[UQU^{*},Q]=0$ for each $U\in \alg{V}\,'$, then $[U,Q]=0$ for each
  $U\in \alg{V}\,'$.
\end{remark}

\begin{proof}  To show that $Q\in \alg{V}(=\alg{V}\,'')$, it will suffice to
  show that $[Q,H]=0$ for any self-adjoint $H\in \alg{V}\,'$ (since
  $\alg{V}\,'$ is a $*$-algebra).  If $H=H^{*} \in \alg{V}\,'$, then
  $U_{t} \equiv e^{itH} \in \alg{V}\,'$ is unitary for all $t\in
  \mathbb{R}$.  By hypothesis, then, $[U_{t}QU_{-t},Q]=0$, for all
  $t\in \mathbb{R}$.
  
  Since $H$ is bounded, $\spectrum{H}$ is a compact subset of
  $\mathbb{R}$.  Consider the one-parameter family $\{ e^{itx} \}
  _{t\in \mathbb{R}}$ of (complex valued) continuous functions on
  $\spectrum{H}$.  Clearly, this family converges uniformly to the
  constant $1$ function as $t\!\rightarrow\!0$.  Employing the
  continuous function calculus~\cite[p. 239]{kad}, it follows that
  $e^{itH}$ converges uniformly to $I$ as $t\!\rightarrow\!0$.  Thus,
  $\lim _{t\rightarrow 0} (U_{t}QU_{-t} )=Q$.  Since $U_{t}QU_{-t}$
  and $Q$ commute, we may write $Q =A_{t}+B_{t}$, $U_{t}QU_{-t}
  =A_{t}+C_{t}$, where $A_{t},B_{t}$ and $C_{t}$ are pairwise
  orthogonal projections.  Then, $0 =\lim _{t\rightarrow 0}
  \norm{U_{t}QU_{-t}-Q} =\lim _{t\rightarrow 0} \norm{B_{t}-C_{t}}$.
  Choose $s>0$ such that $\norm{B_{t}-C_{t}}<\frac{1}{2}$ for all
  $t<s$.  Suppose that $B_{t} \neq 0$ for some $t<s$.  Then
  $\range{B_{t}}\neq \{ 0 \}$ and we may choose a unit vector $x\in
  \range{B_{t}}$.  But then $\norm{(B_{t}-C_{t})x}=\norm{x}=1$, which
  contradicts the fact that $\norm{B_{t}-C_{t}}<\frac{1}{2}$.  Thus,
  $B_{t}=0$ for all $t<s$, and by symmetry $C_{t}=0$ for all $t<s$.
  Hence, for all $t<s$, $U_{t}QU_{-t}=A_{t}=Q$, i.e. $[U_{t},Q]=0$.
  
  Employing the functional calculus for $\spectrum{H}$ again, we see
  that $t^{-1}(e^{itx}-1)$ converges uniformly to $ix$ as
  $t\!\rightarrow\!0$; thus, $t^{-1}(e^{itH}-I)\rightarrow iH$
  uniformly as $t\!\rightarrow\!0$.  We may then compute,
\begin{align}
  (-i)(iH)Q&=\:-i\,\biggl[\lim _{t\rightarrow
    0}(t^{-1}(U_{t}-I))\biggr]Q \:
  =\:-i\,\biggl[\lim _{t\rightarrow 0}(t^{-1}(U_{t}Q-Q))\biggr]  \\
  &=-i\,\biggl[\lim _{t\rightarrow 0}(t^{-1} (QU_{t}-Q))\biggr]
  \:=\:-iQ\,\biggl[\lim _{t\rightarrow 0}(t^{-1} (U_{t}-I))\biggr]
  \\
  &=-iQ(iH). \end{align} The second (and fourth) equalities follow
since right (and left) multiplication by $Q$ is norm continuous.  The
third equality follows since there is an $s>0$ such that $[U_{t},Q]=0$
for all $t<s$.  Therefore, $[H,Q]=0$. \end{proof}

\begin{remark} As before, let $\hil{K}\equiv \range{K}$.  Consider the family of
  subspaces $\hil{Y}$ of $\hil{H}$ such that $\hil{Y}$ contains
  $\hil{K}$ and $\hil{Y}$ is invariant under each element of
  $\alg{R}$.  Since this family is closed under intersection, it will
  contain a unique smallest element which we denote by $\hil{S}$.  It
  is not difficult to see then that $\hil{S}=[\alg{R}''\hil{K}]$.
  (Note that since $\alg{R}$ is (trivially) closed under taking
  adjoints, it follows that $\vn{\alg{R}}=\alg{R}''$.)  Indeed,
  $[\alg{R}''\hil{K}]$ contains $\hil{K}$ and is invariant under each
  element of $\alg{R}$.  Thus, $\hil{S}\subseteq [\alg{R}''\hil{K}]$.
  Conversely, $[\alg{R}''\hil{K}]$ is the smallest subspace of
  $\hil{H}$ that contains $\hil{K}$ and that is invariant under each
  element in $\alg{R}''$.  However, $\hil{S}$ contains $\hil{K}$ and
  $\hil{S}$ is invariant under each element in $\alg{R}''$ (since
  $P_{\hil{S}}\in \alg{R}'=\alg{R}'''$).  Therefore,
  $\hil{S}=[\alg{R}''\hil{K}]$. \label{subspace} \end{remark}

\begin{thm} Let  $\hil{S}$ be the smallest subspace of $\hil{H}$ such
  that $\hil{S}$ contains $\range{K}$ and $\hil{S}$ is invariant under
  $\alg{R}$ (so $\hil{S}=[\alg{R}''\hil{K}]$).  Then, every maximal
  $\alg{R}$-beable algebra for $K$ has the form
  $\alg{L}(\hil{S}^{\perp})\oplus \alg{M}$, where
  $\vn{\alg{R}}P_{\hil{S}}\,\subseteq\,\alg{M}\,\subseteq\,\vn{\alg{R},K}P_{\hil{S}}$,
  and $\alg{M}$ is maximal abelian in $\vn{\alg{R},K}P_{\hil{S}}$.
  \label{rbeable} \end{thm}

\noindent For convenience, we call algebras of the above form
\emph{MRB-algebras} for $K$.  (With this proof in hand, we can
justifiably call these algebras maximal $\alg{R}$-beable algebras.)

\begin{proof} The proof proceeds in two stages.  First, we show that every
  MRB-algebra for $K$ is, in fact, a maximal $\alg{R}$-beable algebra
  for $K$ ((i) below).  Second, we show that every $\alg{R}$-beable
  algebra for $K$ is contained in some MRB-algebra for $K$ ((ii)
  below).
  
  (i) Suppose that $\alg{B}$ is an MRB-algebra for $K$.  That is
  $\alg{B}=\alg{L}(\hil{S}^{\perp})\oplus \alg{M}$, where $\alg{M}$ is
  a maximal abelian subalgebra of $\vn{\alg{R},K}P_{\hil{S}}$, and
  $\vn{\alg{R}}P_{\hil{S}} \subseteq \alg{M}$.
  
  ($\alg{R}$-Priv) Since $\alg{R}$ leaves $\hil{S}$ invariant and
  $\alg{R}P_{\hil{S}}\subseteq \vn{\alg{R}}P_{\hil{S}}\subseteq
  \alg{M}$, it follows that $\alg{R}\subseteq \alg{B}$.
  
  (Beable) Let $\hil{T}\equiv [\alg{B}\hil{K}]$.  By construction of
  $\hil{S}$, $\hil{S}$ contains $\hil{K}$ and is invariant under each
  element of $\alg{B}$.  However, $\hil{T}$ is the smallest subspace
  of $\hil{H}$ that contains $\hil{K}$ and is invariant under each
  element in $\alg{B}$.  Hence, $\hil{T}\subseteq \hil{S}$.
  Conversely, $\hil{T}$ is invariant under each element of $\alg{R}$
  since $\alg{R}$ is contained in $\alg{B}$.  Thus, $\hil{S}=\hil{T}$
  and $\alg{B}=\alg{L}(\hil{T}^{\perp})\oplus \alg{M}$.  But since
  $\alg{M}$ is an abelian subalgebra of $\vn{\alg{R},K}P_{\hil{S}}$,
  it is an abelian subalgebra of $\alg{L}(\hil{T})$.  By
  Theorem~\ref{thm_def}.(i), $\alg{B}$ is beable for $K$.
  
  (Def) We show first that $P_{\hil{S}}$ is in the center of
  $\vn{\alg{R},K}$.  Since $\alg{R}\cup \{ K\}$ is a self-adjoint set,
  $\vn{\alg{R},K}=(\alg{R}\cup \{ K\} )''$.  Let $A\in (\alg{R}\cup \{
  K\} )'=\alg{R}'\cap \{ K\} '$, and let $Bx$ be a generator of
  $\hil{S}$.  (That is, $x\in \hil{K}$ and $B\in \alg{R}''$.)  Since
  $A$ commutes with $K$, $A$ leaves $\hil{K}$ invariant.  Further,
  $[A,B]=0$ since $A\in \alg{R}'=\alg{R}'''$.  Thus, $A(Bx)=B(Ax)\in
  [\alg{R}''\hil{K}]\equiv \hil{S}$, and we may conclude (by linearity
  and continuity of $A$) that $A(\hil{S})\subseteq \hil{S}$.  Since
  the same argument applies to $A^{*}$ (which is also contained in
  $\alg{R}'\cap \{ K\} '$), $\hil{S}$ reduces $A$; and thus,
  $P_{\hil{S}}\in (\alg{R}\cup \{ K\} )''$.  On the other hand,
  $P_{\hil{S}}$ is clearly contained in $(\alg{R}\cup \{ K\} )'$ since
  $\hil{S}$ is invariant under the action of $K$ and under the action
  of the self-adjoint set $\alg{R}$.
  
  Let $U\in (\alg{R} \cup \{ K \} )'=\alg{R}' \cap \{ K \} '$.  Since
  $P_{\hil{S}}\in (\alg{R}\cup \{ K\} )''$, it follows that
  $UP_{\hil{S}}=P_{\hil{S}}U$.  Now let $B\in \alg{B}$.  Then,
  \begin{equation} B=(I-P_{\hil{S}})B(I-P_{\hil{S}})\oplus
    P_{\hil{S}}BP_{\hil{S}},
  \end{equation} where $P_{\hil{S}}BP_{\hil{S}}\in
  \vn{\alg{R},K}P_{\hil{S}}$.  Since $\hil{S}$ reduces $U$, the
  spatial isomorphism $\Phi$ induced by $U$ factors into $\Phi _{1}$,
  the spatial automorphism on $\alg{L}(\hil{S}^{\perp})$ induced by
  $(I-P_{\hil{S}})U(I-P_{\hil{S}})$, and $\Phi _{2}$, the spatial
  automorphism on $\alg{L}(\hil{S})$ induced by
  $P_{\hil{S}}UP_{\hil{S}}$.  Hence,
\begin{equation}
\Phi (B)=\Phi _{1}((I-P_{\hil{S}})B(I-P_{\hil{S}}))\oplus \Phi _{2}(P_{\hil{S}}BP_{\hil{S}}).
\end{equation}
Trivially, $\Phi _{1}((I-P_{\hil{S}})B(I-P_{\hil{S}}))\in
\alg{L}(\hil{S}^{\perp})$.  Furthermore, since $\alg{M}$ is a subset
of $\vn{\alg{R},K}P_{\hil{S}}$, it follows that $\Phi _{2}$ is the
identity automorphism on $\alg{M}$.  To see this, note that,
\begin{equation} P_{\hil{S}}UP_{\hil{S}}\:\in\:P_{\hil{S}}(\alg{R}\cup \{ K\} )'P_{\hil{S}}\:
=\:[\vn{\alg{R},K}P_{\hil{S}}]\,' , \end{equation} where
the final equality follows from~\cite[Proposition 5.5.6]{kad} and the fact
that $P_{\hil{S}}\in (\alg{R}\cup \{ K\} )'$.  Thus, $P_{\hil{S}}UP_{\hil{S}}$ commutes
with every operator in $\vn{\alg{R},K}P_{\hil{S}}$, and $\Phi_{2}$
is the identity automorphism on $\vn{\alg{R},K}P_{\hil{S}}$.  It then follows that
\begin{align}
  \Phi (B)&=\Phi _{1}((I-P_{\hil{S}})B(I-P_{\hil{S}}))\oplus \Phi _{2}(P_{\hil{S}}BP_{\hil{S}}) \\
  &=\Phi _{1}((I-P_{\hil{S}})B(I-P_{\hil{S}}))\oplus
  P_{\hil{S}}BP_{\hil{S}} ,\end{align} which is obviously contained in
$\alg{B}$.  Since $B$ was an arbitrary element of $\alg{B}$, it
follows that $\Phi (\alg{B})\subseteq \alg{B}$.  Moreover, this map is
onto, for given $A\in \alg{B}$,
\begin{align} \Phi ((I-P_{\hil{S}})\Phi ^{-1}(A)(I-P_{\hil{S}})\oplus P_{\hil{S}}AP_{\hil{S}})
  &= (I-P_{\hil{S}})A(I-P_{\hil{S}})\oplus
  \Phi _{2}(P_{\hil{S}}AP_{\hil{S}}) \\
  &=(I-P_{\hil{S}})A(I-P_{\hil{S}})\oplus
  P_{\hil{S}}AP_{\hil{S}}\;=\;A. \end{align} Therefore, $\Phi
(\alg{B})=\alg{B}$.

(Maximality) To see that $\alg{B}$ is a maximal $\alg{R}$-beable
algebra for $K$, it suffices to show that (a) every $\alg{R}$-beable
algebra for $K$ is contained in an MRB-algebra for $K$, and (b) if
$\alg{B}_{1}$ and $\alg{B}_{2}$ are distinct MRB-algebras for $K$,
then $\alg{B}_{1} \not\subseteq \alg{B}_{2}$.  We establish (a) in
(ii) below.  For (b) it suffices to note that if $\alg{B}_{1}$ and
$\alg{B}_{2}$ are distinct MRB-algebras for $K$, then $\alg{B}_{1}
=\alg{L}(\hil{S}^{\perp})\oplus \alg{M}_{1}$ and $\alg{B}_{2}
=\alg{L}(\hil{S}^{\perp})\oplus \alg{M}_{2}$, where $\alg{M}_{1}$ and
$\alg{M}_{2}$ are distinct maximal abelian subalgebras of
$\vn{\alg{R},K}P_{\hil{S}}$ (each containing
$\vn{\alg{R}}P_{\hil{S}}$).  Thus, $\alg{M}_{1}\not\subseteq
\alg{M}_{2}$ and $\alg{B}_{1}\not\subseteq \alg{B}_{2}$.

(ii) Suppose that $\alg{B}$ is $\alg{R}$-beable for $K$.  Since
$\alg{B}\subseteq \alg{B}^{-}$, it will suffice to show that
$\alg{B}^{-}$ is contained in an MRB-algebra for $K$ because, by
Lemma~\ref{juji}, $\alg{B}^{-}$ is $\alg{R}$-beable for $K$.  Thus, we
may assume that $\alg{B}$ is a von Neumann algebra.
 
Once again, let $\hil{T}\equiv [\alg{B}\hil{K}]$.  Obviously,
$\hil{T}$ reduces $\alg{B}$, and $P_{\hil{T}}\in \alg{B}'$.  Since
$\alg{B}$ is a von Neumann algebra $\alg{B}P_{\hil{T}}$ is a von
Neumann algebra acting on $\hil{T}$~\cite[Proposition 5.5.6]{kad}.
Likewise, $\alg{B}(I-P_{\hil{T}})$ is a von Neumann algebra acting on
$\hil{T}^{\perp}$.  Let $\alg{M}\equiv \alg{B}P_{\hil{T}}$.  Then we
have $\alg{B}=\alg{B}(I-P_{\hil{T}})\oplus \alg{M}$, where each
summand is a von Neumann algebra.  Since $\alg{B}$ is beable for $K$,
$\alg{M}$ is in fact an abelian subalgebra of
$\alg{L}(\hil{T})$~(Theorem~\ref{thm_def}.(i)).  We show that
$\hil{T}=\hil{S}$ and that $\alg{M}\subseteq
\vn{\alg{R},K}P_{\hil{S}}$.  (Clearly, once $\hil{T}=\hil{S}$ has been
established, we will automatically have $\vn{\alg{R}}P_{\hil{S}}
\subseteq \alg{M}$, since $\vn{\alg{R}}\subseteq \alg{B}$ and
$\vn{\alg{R}}$ leaves $\hil{S}$ invariant.)

$\hil{S}$ is clearly a subspace of $\hil{T}$ since $\alg{R} \subseteq
\alg{B}$.  In order to show that $\hil{T} \subseteq \hil{S}$, let
$\overline{F}$ be a projection in $\alg{B}$.  Since $\hil{T}$ reduces
$\overline{F}$, $\overline{F}=F_{0}\oplus F\in
\alg{L}(\hil{T}^{\perp})\oplus \alg{M}$.  Choose $\theta \in
\mathbb{R}$ such that $e^{-i\theta }\neq \pm 1$.  Let $\overline{U}
=P_{T^{\perp}} \oplus P_{\hil{S}}\oplus (e^{i\theta }P_{T\wedge
  S^{\perp}})$, and let $U=P_{\hil{S}}\oplus (e^{i\theta }P_{T\wedge
  S^{\perp}})\in \alg{L}(\hil {T})$.  Since $\alg{R}\subseteq
\alg{B}$, $\alg{R}$ leaves $\hil{T}$ invariant, and (by construction)
$\alg{R}$ leaves $\hil{S}$ invariant.  Thus, $\alg{R}$ leaves $\hil{S}
^{\perp} \wedge \hil{T}$ invariant, and $\overline{U}\in \alg{R}'$.
Furthermore, $\hil{K} \subseteq \hil{S}$, and $\overline{U}\!\!\mid
_{S}=I$.  Thus, $\overline{U}\in \{ K \} '$.  Since $\overline{U}\in
(\alg{R}'\cap \{ K \} ')$, it follows by (Def) that
$\overline{U}\alg{B}\overline{U}^{*}=\alg{B}$; and since $\hil{T}$
reduces $\overline{U}$, it also follows that $U\alg{M}U^{*}=\alg{M}$.
In particular, both $F$ and $UFU^{*}$ are in the abelian algebra
$\alg{M}$.

Since $F$ and $UFU^{*}$ commute, there are mutually orthogonal
projections $A,B,C$ on $\hil{T}$ such that $F =A+B$, $UFU^{*} =A+C$.
To see that $B=0$, let $v \in \range{B}\subseteq \hil{T}$. Then,
\begin{align}
  Fv &=v, \label{eqn_three} \\
  UFU^{*}v &=0. \label{eqn_four} \\
  \intertext{Now, we may also write $v=w+w'$ (uniquely), where $w\in
    \hil{S}$ and $w' \in (\hil{S} ^{\perp } \wedge \hil{T} )$.
    Using~(\ref{eqn_four}) (and $U^{-1}(0)=0$), we get:}
  FU^{*} (w+w') &=0. \\
  \intertext{But, by the definition of $U$ (and using $v=w+w'$), this
    implies that}
  F(w+e^{-i\theta } w') &=0 \label{eqn_five} \\
  \intertext{and}
  F(v-w'+e^{-i\theta } w') &=0. \\
  \intertext{Next, using~(\ref{eqn_three}) and $e^{-i\theta}\neq 1$}
  Fw' &=(1-e^{-i \theta })^{-1}v, \label{eqn_six} \\
  \intertext{thus,}
  UFU^{*} Fw' &=(1-e^ {-i\theta })^{-1} UFU^{*}v, \\
  \intertext{and we can see by~(\ref{eqn_four}) that this last
    expression vanishes.  However, $UFU^{*}$ and $F$ commute on
    $w'(\in \hil{T})$.  It follows that $F(UFU^{*} ) w'=0$ as well.
    We can then compute, using the definition of $U$ and $e^{-i\theta
      }\neq 0$,}
  FUFw' &=0. \\
  \intertext{By~(\ref{eqn_six}),}
  FU[(1-e^{-i\theta } )^{-1} v] &=0.  \\
  \intertext{But since $(1-e^{-i\theta})^{-1}\neq 0$,}
  FUv &=0,  \\
  FU(w+w') &=0, \\
  F(w+e^{i\theta } w') &=0, \label{eqn_seven} \\
  \intertext{again using the definition of $U$ in the move
    to~(\ref{eqn_seven}).  But now~(\ref{eqn_five})
    and~(\ref{eqn_seven}) together entail:}
  F(w+e^ {-i\theta }w'-(w+e^{i\theta } w')) &=0, \\
  F((e^{-i\theta } -e^{i\theta } )w') &=0. \label{extra} \\
  \intertext{However, $e^{-i\theta } -e^{i\theta } \neq 0$ (since
    $e^{-i\theta }\neq \pm 1$).  Thus, by~(\ref{eqn_seven})
    and~(\ref{extra}),}
  Fw &=0, \\
  Fw' &=0, \\
  Fv &=F(w+w') \\
  &=0. \end{align} Thus, $Fv=0$.  But, by~(\ref{eqn_three}), $Fv=v$.
Hence, $v=0$ and $B=0$.

Now we may repeat a similar argument with $F$ replaced by $UFU^{*}$,
and $U^{*}(UFU^{*})U=F$.  (The only change to the argument is that,
throughout, $\theta$ must be interchanged with $-\theta$, since $U$ is
interchanged with $U^{*}$).  It follows that $C=0$ as well, and thus
$UFU^{*} =F$.

We chose $U$, however, so that if $UF=FU$, then $P_{\hil{S}}
F=FP_{\hil{S}}$.  Indeed, a routine calculation shows that
$P_{\hil{S}}=(e^{i \theta }-1)^{-1}[e^{i\theta } P_{\hil{T}}-U]$.
Furthermore,
\begin{align}
  \overline{F}P_{\hil{S}} &=[(I-P_{\hil{T}})\overline{F}
  (I-P_{\hil{T}} ) +P_{\hil{T}}\overline{F}P_{\hil{T}}] P_{\hil{S}}
  \\
  &=[(I-P_{\hil{T}})\overline{F} (I-P_{\hil{T}} ) +F] P_{\hil{S}} \\
  &=FP_{\hil{S}}=P_{\hil{S}} F=P_{\hil{S}} \overline{F}
\end{align}
since $P_{\hil{S}} (I-P_{\hil{T}})=0$.  Thus,
$\overline{F}P_{\hil{S}}=P_{\hil{S}}\overline{F}$, for all projections
$\overline{F} \in \alg{B}$.  Since $\alg{B}$ is (by hypothesis) a von
Neumann algebra, each $A \in \alg{B}$ is a norm-limit of linear
combinations of projections in $\alg{B}$.  Thus, $AP_{\hil{S}}
=P_{\hil{S}}A$, for all $A\in \alg{B}$ and $\hil{S}$ reduces
$\alg{B}$.  Since $\hil{K} \subseteq \hil{S}$, and since $\hil{T}$ is
the smallest subspace that contains $\hil{K}$ and reduces $\alg{B}$,
it follows that $\hil{T}\subseteq \hil{S}$.

We have now shown that $\hil{T}=\hil{S}$ and that, accordingly,
$\alg{M}$ is an abelian von Neumann subalgebra of $\alg{L}(\hil{S})$.
All that remains is to show that $\alg{M}\subseteq
\vn{\alg{R},K}P_{\hil{S}}$.  Since $\alg{M}$ is a von Neumann algebra,
it will suffice to show that for any projection $Q\in \alg{M}$,
$Q\in\vn{\alg{R},K}P_{\hil{S}}$.  Let $U\in (\alg{R} ' \cap \{ K \}
')$.  Then, by (Def), $U\alg{B}U^{*}=\alg{B}$.  Since $U$ is reduced
by $\hil{S}=[\alg{R}''\hil{K}]$, $U\alg{M}U^{*}=\alg{M}$.  In
particular, $Q,UQU^{*}\in \alg{M}$.  Since $\alg{M}$ is abelian,
$UQU^{*}Q=QUQU^{*}$.  But $U$ was an arbitrary unitary operator in
$\alg{R}'\cap \{ K\}'$.  Applying Lemma~\ref{lemma_stone}, with
$\alg{V}=\vn{\alg{R},K}$, we may conclude that $Q\in \vn{\alg{R},K}$.
Moreover, $QP_{\hil{S}}=Q$, since $Q\leq P_{\hil{S}}$.  Thus, $Q\in
\vn{\alg{R},K}P_{\hil{S}}$ and $\alg{B}$ is contained in an
MRB-algebra for $K$. \end{proof}

Theorem~\ref{rbeable} shows that the requirement that $\alg{B}$ be
maximal $\alg{R}$-beable places a significant restriction on the
structure of $\alg{B}$.  However, it still does not follow that there
is always a unique maximal $\alg{R}$-beable algebra for $K$.

\begin{example}  Let $\hil{H}$ be three-dimensional, 
  choose an orthonormal basis $\{ r_{1},r_{2},r_{3} \}$, and let
  $\alg{R}=\{ R\}$, where $R$ is a self-adjoint operator on $\hil{H}$
  with only two eigenvalues and corresponding eigenspaces, $[r_{1}]$
  and $[r_{2},r_{3}]$.  Choose another orthonormal basis $\{
  w_{1},w_{2},w_{3} \}$ so that the vectors $w_{1}$ and $w_{2}$ do not
  lie inside either of $R$'s eigenspaces, and $P_{[r_{2},r_{3}]}w_{1}$
  and $P_{[r_{2},r_{3}]}w_{2}$---the orthogonal projections of $w_{1}$
  and $w_{2}$ onto the plane $[r_{2},r_{3}]$---are neither parallel
  nor orthogonal.  Let $K$ be any positive, trace-1 operator with
  three distinct nonzero eigenvalues corresponding to the eigenspaces
  $[w_{1}]$, $[w_{2}]$, and $[w_{3}]$.  By construction, $\{
  P_{[r_{1}]}w_{1},P_{[r_{2},r_{3}]}w_{1},P_{[r_{2},r_{3}]}w_{2}\}$
  spans $\hil{H}$, and thus $\hil{S}=[\alg{R}''\hil{K}]=\hil{H}$.  It
  follows from Theorem~\ref{rbeable} that any maximal abelian
  subalgebra of $\vn{R,K}$ containing $R$ is maximal $R$-beable for
  $K$.  In fact, $\vn{R,K}$ contains \emph{two} such subalgebras.
    
  Let $A_{1}$ be any nondegenerate, self-adjoint operator with
  one-dimensional (mutually orthogonal) eigenspaces
  $[P_{[r_{2},r_{3}]}w_{1}]$, $[P_{[r_{2},r_{3}]}w_{1}]^{\perp}\wedge
  [r_{2},r_{3}]$, and $[r_{1}]$.  Let $A_{2}$ be any nondegenerate,
  self-adjoint operator with one-dimensional eigenspaces
  $[P_{[r_{2},r_{3}]}w_{2}], [P_{[r_{2},r_{3}]}w_{2}]^{\perp}\wedge
  [r_{2},r_{3}]$ and $[r_{1}]$.  Since $\vn{R,K}$ contains the
  spectral projections of both $R$ and $K$, and the projections in
  $\vn{R,K}$ form an ortholattice, the projections onto all the
  eigenspaces of each $A_{i}$ lie in $\vn{R,K}$ (for example,
  $[P_{[r_{2},r_{3}]}w_{1}]$ may be expressed as $([w_{1}]\vee
  [r_{1}])\wedge [r_{2},r_{3}]$, and similarly for
  $[P_{[r_{2},r_{3}]}w_{2}]$).  It follows that each $A_{i}\in
  \vn{R,K}$.  Let $\vn{A_{i}}$ be the von Neumann algebra generated by
  $A_{i}$.  Since the projections onto $[P_{[r_{2},r_{3}]}w_{1}]$ and
  $[P_{[r_{2},r_{3}]}w_{2}]$ fail to commute (by construction of
  $w_{1}$ and $w_{2}$), $A_{1}$ and $A_{2}$ do not commute and
  $\vn{A_{1}}$ and $\vn{A_{2}}$ are distinct. And, since each $A_{i}$
  is nondegenerate, each $\vn{A_{i}}$ is maximal abelian in $\bh$, and
  thus maximal abelian in $\vn{R,K}$.  Moreover, $R\in \vn{A_{i}}$
  since $R$ commutes with each $A_{i}$.  Therefore, $\vn{A_{1}}$ and
  $\vn{A_{2}}$ are distinct maximal $R$-beable algebras for $K$.
\end{example}

Although the above example shows that we cannot always expect there to
be a unique maximal $\alg{R}$-beable algebra for $K$, there are at
least two important cases where uniqueness does hold:

\begin{cor} \ \  
\begin{enumerate} 
\item If $K\in \alg{R}'$, then the \emph{unique} maximal
  $\alg{R}$-beable algebra for $K$ is $\alg{L}(\hil{S}^{\perp})\oplus
  \vn{\alg{R},K}P_{\hil{S}} $.
\item If $K=P_{v}$, for some $v\in \hil{H}$, then the \emph{unique}
  maximal $\alg{R}$-beable algebra for~$K$~is
  $\alg{L}(\hil{S}^{\perp})\oplus \vn{\alg{R}}P_{\hil{S}}$.
  \end{enumerate}
\label{unique} \end{cor}

\begin{proof} (i) Since elements of $\alg{R}$ pairwise commute, and
  $K\in \alg{R}'$, it follows that $\vn{\alg{R},K}$ is abelian, as is
  $\vn{\alg{R},K}P_{\hil{S}}$.  Therefore,
  $\alg{L}(\hil{S}^{\perp})\oplus \vn{\alg{R},K}P_{\hil{S}}$ is itself
  the unique maximal $\alg{R}$-beable algebra for $K$.
  
  (ii) Recall that $\hil{S}=[\alg{R}''\hil{K}]$.  Thus, in this case,
  $\hil{S}=[\alg{R}''v]$.  Since $\vn{\alg{R}}P_{\hil{S}}$ is abelian
  and $v$ is a cyclic vector for $\vn{\alg{R}}P_{\hil{S}}$, it follows
  that $\vn{\alg{R}}P_{\hil{S}}$ is maximal abelian as a subalgebra of
  $\alg{L}(\hil{S})$~\cite[Corollary 7.2.16]{kad}.  Accordingly,
  $\vn{\alg{R}}P_{\hil{S}}$ is the unique maximal abelian subalgebra
  $\alg{M}$ of $\vn{\alg{R},K}P_{\hil{S}}$ with the property that
  $\vn{\alg{R}}P_{\hil{S}}\subseteq \alg{M}$.  \end{proof}

\begin{remark} When $\alg{R}=\{ K\}$, case (i) applies, and
  the maximal $\alg{R}$-beable subalgebra consists of exactly those
  observables that share with $K$ the spectral projections that
  project onto $K$'s range.  This set of observables are those taken
  to be determinate in most `modal' interpretations of quantum
  theory~\cite{clifmod,clifmod2,zimba}.  On the other hand, case (ii)
  strengthens and generalizes (to observables with continuous spectra)
  the theorem proved in~\cite{bubclif}, which is the basis for the
  alternative modal interpretation of quantum theory developed by
  Bub~\cite{bub}.\end{remark}

In what follows, we will denote the von Neumann algebra referred to in
Corollary~\ref{unique}.(ii) by $\alg{B}(\alg{R},v)$.  That is,
$\alg{B}(\alg{R},v)=\alg{L}(\hil{S}^{\perp})\oplus
\vn{\alg{R}}P_{\hil{S}}$, where $\hil{S}=[\alg{R}''v]$.  We end this
section with two applications of Corollary~\ref{unique}.(ii) to the
Copenhagen interpretation of quantum theory that are facilitated by
the following more tractable characterization:

\begin{prop} Let $A$ be in $\bh _{\mathrm{sa}}$.  \begin{enumerate} 
  \item If $A\in \alg{R}'$ and $Av\in [\alg{R}''v]$, then $A\in
    \alg{B}(\alg{R},v)$.
  \item If $A$ does not leave $[\alg{R}''v]$ invariant, then $A\not\in
    \alg{B}(\alg{R},v)$.
\end{enumerate} \label{the_test} \end{prop}

\begin{proof} (i) Suppose that $A\in \alg{R}'=\alg{R}'''$ and that
  $Av\in [\alg{R}''v]$.  Using Lemma~\ref{simplify} for the
  $C^{*}$-algebra $\alg{R}''$, it follows that $A$ leaves
  $[\alg{R}''v]=\hil{S}$ invariant.  Since $A$ is self-adjoint, $A$
  also leaves $\hil{S}^{\perp}$ invariant and $A\in
  \alg{L}(\hil{S}^{\perp})\oplus \alg{L}(\hil{S})$.
  
  Since $P_{\hil{S}}\in \vn{\alg{R}}'$, the commutant of
  $\vn{\alg{R}}P_{\hil{S}}$ relative to $\alg{L}(\hil{S})$ is
  $P_{\hil{S}}\vn{\alg{R}}'P_{\hil{S}}=P_{\hil{S}}\alg{R}'P_{\hil{S}}$~\cite[Prop.
  5.5.6]{kad}.  Clearly, then, $P_{\hil{S}}AP_{\hil{S}}$ is in the
  commutant of $\vn{\alg{R}}P_{\hil{S}}$.  However, since
  $\vn{\alg{R}}P_{\hil{S}}$ is maximal abelian,
  $P_{\hil{S}}AP_{\hil{S}}\in \vn{\alg{R}}P_{\hil{S}}$.  Therefore,
  $A\in \alg{L}(\hil{S}^{\perp})\oplus
  \vn{\alg{R}}P_{\hil{S}}=\alg{B}(\alg{R},v)$.
  
  (ii) is trivial, since each element in $\alg{B}(\alg{R},v)$ leaves
  $[\alg{R}''v]$ invariant.  \end{proof}
  
One of Bell's motivations for distinguishing beables from observables
was that the distinction makes ``\ldots explicit some notions already
implicit in, and basic to, ordinary quantum theory.  For, in the words
of Bohr, `it is decisive to recognize that, however far the phenomena
transcend the scope of classical physical explanation, the account of
all evidence must be expressed in classical terms'.  It is the
ambition of the theory of local beables to bring these `classical
terms' into the equations, and not relegate them entirely to the
surrounding talk''~\cite[p. 52]{bell3}.  One can fulfil this ambition
by understanding Bohr's assertions, about the possibility of
attributing certain observables determinate values in certain
measurement contexts, as arising from selecting the maximal set of
observables that can be determinate together with the determinacy of
whatever measurement results are actually obtained in a given
measurement context.  Thus we propose to understand the Copenhagen
interpretation, \emph{not} as relying on a collapse of the state
vector when a measurement occurs, but rather as selecting for beable
status the maximal $R$-beable subalgebra determined by the
``privileged'' pointer observable $R$ of the measuring system and the
pure entangled state $P_{v}$ of the composite measured/measuring
system (see ~\cite{bub5,howard} for related proposals that do not
adopt an algebraic approach).  It then becomes possible to make
precise the (hitherto obscure) sense in which a measurement, for Bohr,
can `make determinate' the observable that was measured, as well as
make determinate certain observables of spacelike-separated systems.

 \begin{example}[Ideal Measurement] Let $\hil{H}$ and $\hil{G}$ be
   separable Hilbert spaces for an apparatus and object respectively,
   let $\overline{R}$ be the apparatus pointer observable on $\hil{H}$
   with eigenvectors $x_{n}$, and let $\overline{M}$ be the measured
   observable on $\hil{G}$ with eigenvectors $y_{n}$ and respective
   eigenvalues $\lambda_{n}$.  Let $R$ be the self-adjoint operator
   $\overline{R}\otimes I$ on $\hil{H}\otimes \hil{G}$ and let $M$ be
   the self-adjoint operator $I\otimes \overline{M}$ on
   $\hil{H}\otimes \hil{G}$.  Note that
   $\vn{R}=\vn{\overline{R}}\otimes I$.
   
   Prior to an entangling measurement interaction that strictly
   correlates the values of $\overline{M}$ with $\overline{R}$, the
   total state will be $v_{0}=x_{0}\otimes\sum c_{n}y_{n}$, where
   $\sum\abs{c_{n}}^{2}=1$ and $x_{0}$ is the `ground state' of the
   pointer observable.  When two or more of the coefficients
   $\{c_{n}\}$ are nonzero, and two or more of the $\{\lambda_{n}\}$
   unequal, then the pre-measurement maximal $R$-beable algebra for
   $P_{v_{0}}$, $\alg{B}(R,v_{0})$, will fail to contain $M$.  For
   every element of $\hil{S}=[\vn{R}v_{0}]$ has the form $x\otimes\sum
   c_{n}y_{n}$ for some $x\in\hil{H}$, yet $Mv_{0}=x_{0}\otimes\sum
   c_{n}\lambda_{n}y_{n}$ which is not of the required form.  Thus $M$
   fails to leave $\hil{S}$ invariant, and $M\not\in \alg{B}(R,v_{0})$
   by Corollary~\ref{the_test}.(ii).
   
   On the other hand, after the unitary evolution that affects the
   measurement, the state is $v=\sum c_{n}(x_{n}\otimes y_{n})$.  If
   $Q_{n}$ is the projection onto the one-dimensional subspace
   $[x_{n}]$ of $\hil{H}$, it follows that $Q_{n}\otimes I\in \vn{R}$.
   We then have, $Mv=\sum c_{n}\lambda_{n}(x_{n}\otimes y_{n}) =\Bigl[
   \sum \lambda_{n}(Q_{n}\otimes I)\Bigr] v$, and $\sum
   \lambda_{n}(Q_{n}\otimes I)\in \vn{R}$.  Since $M$ commutes with
   $R$, both conditions of Corollary~\ref{the_test}.(i) are satisfied,
   and $M\in \alg{B}(R,v)$!  Thus the act of measuring $M$ has, in a
   sense, \emph{made} $M$ determinate, but not via any physical
   disturbance (cf. Bohr's~\cite[p. 317]{bohr5} well-known and
   oft-repeated caution against speaking of `creation of physical
   attributes of objects by measurements').  Rather, both before and
   after the measurement one constructs the maximal set of observables
   that, together with the pointer observable $R$, can have
   simultaneously determinate values, and these purely formal
   constructions, designed to secure a maximally complete account of
   each stage of the measurement process in classical terms, forces
   one to different verdicts concerning the determinacy of $M$.
\end{example}

\begin{example}[EPR Correlations---Spin Case] Let $\hil{H}_{1}$ and
  $\hil{H}_{2}$ be two-dimensional Hilbert spaces and let $\sigma
  _{xi},\sigma _{yi},\sigma _{zi}$ be the Pauli spin operators on
  $\hil{H}_{i}$, for $i=1,2$.  For convenience, we write vectors in
  Dirac's ket notation and suppress tensor products between vectors;
  for example, $\ket{\sigma _{x1}=+1}\ket{\sigma _{x2}=-1}$ denotes an
  eigenvector of $\sigma _{x1}\otimes\sigma _{x2}$ with eigenvalue
  $-1$.  Let $\ket{\mathsf{s}}$ be the singlet state in
  $\hil{H}_{1}\otimes \hil{H}_{2}$ which, expanded in the basis of
  eigenvectors for $\sigma _{x1}$, is \begin{equation}
    \label{singlet!}  \ket{\mathsf{s}}=\frac{1}{\sqrt{2}}\Bigl(
    \ket{\sigma _{x1}=+1}\ket{\sigma _{x2}=-1}-\ket{\sigma
      _{x1}=-1}\ket{\sigma _{x2}=+1}\Bigr) . \end{equation} This state
  also assumes the same form relative to the $y$- and $z$-bases, thus
  it predicts that identical spin components of the two particles will
  always be found on measurement to be anti-correlated long after the 
  particles have interacted and separated.  Exploiting
  correlations of the exact same kind between the positions and
  momenta of two particles (whose analysis we defer until the next
  section), Einstein, Podolsky, and Rosen~\cite{epr} argued for the
  joint determinacy of incompatible observables on the basis of the
  following `reality' criterion: ``\emph{If, without in any way
    disturbing a system, we can predict with certainty (i.e., with
    probability equal to unity) the value of a physical quantity, then
    there exists an element of physical reality corresponding to this
    physical quantity}" \cite[p.  777]{epr}.  In the case of
  incompatible spin components in the state~(\ref{singlet!}), EPR's
  argument is straightforward.  If $\sigma_{x1}$ were measured, then,
  regardless of the value obtained, $\sigma_{x2}$'s value could be
  predicted with certainty without in any way disturbing particle $2$, 
  spacelike-separated from $1$.
  It then would follow from the reality criterion that $\sigma_{x2}$
  has a value (is an `element of reality') quite apart from whether
  $\sigma_{x1}$ is actually measured on particle $1$.  But then, by exactly parallel reasoning from
  the possibility of measuring $\sigma_{y1}$, $\sigma_{y2}$ must have
  a value as well---and yet it fails to commute (or, indeed, share any
  eigenvectors) with $\sigma_{x2}$.  Bohr's response to EPR's argument
  pointed to an ambiguity in their phrase ``without in any way
  disturbing a system'': ``Of course there is in a case like that just
  considered no question of a mechanical disturbance of the
  system\ldots.[but] there is essentially the question of \emph{an
    influence on the very conditions which define the possible types
    of predictions regarding the future behaviour of the
    system}"~\cite[p. 148]{bohr}.  The phrase Bohr italicizes here has seemed
  opaque to many commentators (not least, Bell~\cite[p. 155]{bell3}).  Yet by
  employing the appropriate maximal $R$-beable subalgebras, it is
  possible to understand how measuring the $x$-spin (respectively, 
  $y$-spin) of particle $1$
  can have an affect on the conditions that permit the ascription of a 
  definite value to the $x$-spin (respectively, $y$-spin) of particle $2$.
  
  Let $\hil{H}_{0}$ be a three-dimensional Hilbert space, and let
  $R_{0}$ be a self-adjoint operator on $\hil{H}_{0}$ whose
  eigenvalues $-1,0$ and $1$ represent the different possible states
  of the pointer observable on an apparatus ready to measure
  $\sigma_{x1}$.  Prior to the measurement, the total state of
  apparatus and particles is
  $\ket{v_{0}}=\ket{R_{0}=0}\ket{\mathsf{s}}$ with $\ket{R_{0}=0}$ the
  apparatus ground state.  As before, take $R\equiv R_{0}\otimes
  I\otimes I$.  Clearly, $[\vn{R}\ket{w}]$ consists only of vectors of
  the form $\ket{t}\ket{\mathsf{s}}$ for some $\ket{t}\in
  \hil{H}_{0}$, since $\vn{R}=\vn{R_{0}}\otimes (I\otimes
  I)=\vn{R_{0}}\otimes I$.  However, $(I\otimes \sigma _{x1}\otimes
  I)\ket{v_{0}}=\ket{R_{0}=0}\ket{u}$, where
\begin{equation}
\ket{u}=\frac{1}{\sqrt{2}} \Bigl( \ket{\sigma _{x1}=+1}
\ket{\sigma _{x2}=-1}+\ket{\sigma
_{x1}=-1}\ket{\sigma _{x2}=+1} \Bigr) , \end{equation}
and $\ket{u}\perp\ket{\mathsf{s}}$.  Thus, by Corollary~\ref{the_test}.(ii),
$I\otimes \sigma
_{x1}\otimes I\not\in \alg{B}(R,v_{0})$.
A similar argument shows that none of $(I\otimes I\otimes \sigma
_{x2}), (I\otimes \sigma _{y1}\otimes I)$, or $(I\otimes I\otimes \sigma
_{y2})$ lie in
$\alg{B}(R,v_{0})$.

However, after the measurement of $\sigma_{x1}$ actually occurs, it
results in the entangled state
\begin{equation} \ket{v}\equiv \frac{1}{\sqrt{2}} \Bigl(
\ket{R_{0}=+1}\ket{\sigma
_{x1}=+1}\ket{\sigma _{x2}=-1}
-\ket{R_{0}=-1}\ket{\sigma _{x1}=-1}\ket{\sigma _{x2}=+1} \Bigr) .
\end{equation}
Now, $I\otimes \sigma _{x1}\otimes \sigma _{x2}$ commutes with $R$,
and $(I\otimes \sigma _{x1}\otimes \sigma _{x2})\ket{v}=-\ket{v}\in
[\vn{R}\ket{v}]$.  Thus, by Corollary~\ref{the_test}.(i), $I\otimes
\sigma _{x1}\otimes \sigma _{x2}\in \alg{B}(R,v)$.  Moreover, it is
easy to see that $(I\otimes \sigma _{x1}\otimes
I)\ket{v}=(R_{0}\otimes I\otimes I)\ket{v}\in [\vn{R}\ket{v}]$. Thus,
$I\otimes \sigma _{x1}\otimes I\in \beable{R}{v}$.  But since $\sigma
_{x1}^{2}=I$,
\begin{equation} (I\otimes \sigma _{x1}\otimes \sigma _{x2})(I\otimes
  \sigma _{x1}\otimes I)=I\otimes I\otimes \sigma _{x2},
\end{equation} and the latter lies in $\alg{B}(R,v)$ as well.  On the
other hand, it is not difficult to show that $I\otimes \sigma
_{y1}\otimes I\not\in \alg{B}(R,v)$.  First, \begin{multline}
  (I\otimes \sigma _{y1}\otimes I)\ket{v}=-i \Bigl(
  \ket{R_{0}=+1}\ket{\sigma
    _{x1}=-1}\ket{\sigma _{x2}=-1} \\
  +\ket{R_{0}=-1}\ket{\sigma _{x1}=+1}\ket{\sigma _{x2}=+1} \Bigr) .
\end{multline}
However, since $\vn{R}=\vn{R_{0}}\otimes I\otimes I$, the generic
element of $[\vn{R}\ket{v}]$ has the form,
\begin{equation} \ket{t}\ket{\sigma _{x1}=+1}\ket{\sigma
_{x2}=-1}-\ket{u}\ket{\sigma _{x1}=-1}\ket{\sigma _{x2}=+1},
\end{equation}
for some $\ket{t},\ket{u}\in \hil{H}_{0}$. Thus, $(I\otimes \sigma
_{y1}\otimes I)\ket{v}\not\in [\vn{R}\ket{v}]$, and it follows from
Corollary~\ref{the_test}.(ii) that $I\otimes \sigma _{y1}\otimes
I\not\in \alg{B}(R,v)$.  A similar argument shows that $I\otimes
I\otimes \sigma _{y2}\not\in \alg{B}(R,v)$.  Thus we see how
\emph{once $\sigma _{x1}$ is actually measured}, both it and $\sigma
_{x2}$ `become determinate' in just the same nonmechanical sense as
explained at the end of the previous example---and this occurs at the
expense of the determinacy of the $y$-spins of the particles.  Of
course, a parallel analysis applies, by symmetry, if $\sigma _{y1}$
were actually measured instead; in that case, it would be legitimate
to ascribe determinacy to both $y$-spins of the particles at the
expense of their $x$-spins.  In \emph{neither} case (i.e. in neither
the $x_{1}$- or $y_{1}$-spin measurement context) does it follow that
\emph{both} $\sigma _{x2}$ and $\sigma _{y2}$ are determinate.  Thus
the EPR argument fails for exactly the reason suggested by the phrase
Bohr sets in italics in the passage cited above.

Finally, it is worth addressing Schr\"{o}dinger's
\cite[Sec.~12]{schro} clever modification of the EPR argument.  In
terms of spin, his proposal was that one consider measuring $\sigma
_{x1}$ at the same time $t$ (in some frame) as $\sigma _{y2}$ is
measured.  The latter measurement allows one to directly ascertain the
value of $\sigma _{y2}$ while the former's measurement result,
obtained at a distance (`without in any way disturbing the system'),
allows one to infer the value of $\sigma _{x2}$ indirectly
via~(\ref{singlet!})'s strict $x$-spin correlation and the EPR reality
criterion.  It would thus appear, not only that $\sigma _{x2}$ and
$\sigma _{y2}$ must by simultaneously determinate at $t$, but can be
simultaneously \emph{known}!  (According to Schr\"{o}dinger, we have
`hypermaximal' knowledge of the state of particle 2.)

Of course, the value of $\sigma_{x2}$ that becomes `known' by such a
procedure will have no predictive significance for a measurement of
$\sigma _{x2}$ that occurs later than $t$, thus the uncertainty
principle is not contradicted.  More importantly, an analysis along
the lines set forth above shows that Schr\"{o}dinger's experiment
cannot be used to contradict the indeterminacy principle for $\sigma
_{x2}$ and $\sigma _{y2}$ either.  Assuming both $\sigma _{x1}$ and
$\sigma _{y2}$ are actually measured at time $t$ in state
~(\ref{singlet!}), and modelling the two measurements in terms of
strict correlations to the values of two pointer observables $R_{1}$
and $R_{2}$, let the final post-measurement state be $\ket{v_{t}}$.
To ascertain which observables can be regarded as determinate in this
measurement context, we must now take our set of preferred observables
to include \emph{both} pointer observables.  It is then easy to show
that $\sigma _{y2}$ (and $\sigma _{x1}$) lies in
$\alg{B}(\{R_{1},R_{2}\},v_{t})$ but \emph{not} $\sigma _{x2}$ (or
$\sigma _{y1}$).  Thus performing a direct measurement of $\sigma
_{y2}$ renders invalid Schr\"{o}dinger's use of the EPR reality
criterion to secure a value for $\sigma _{x2}$ in the given
measurement context.  It follows that the EPR reality criterion cannot
be part of the Copenhagen interpretation (if our reconstruction of the
interpretation is correct), but is valid only in the special case
where there is no direct measurement being made of observables
incompatible with ones whose values are predictable with certainty on
the basis of the criterion.
\end{example}

\subsection{$\alg{R}$-beable algebras for arbitrary pure states.}
We now discuss the extension of Theorem~\ref{rbeable} to the case of
an arbitrary (not necessarily normal) pure state on a $C^{*}$-algebra
$\alg{A}$ (either abstract or concrete).  Although our results are
limited, they still permit a characterization of Bohr's response to
the \emph{original} EPR argument, which in fact employed a singular
state of two particles with strictly correlated positions and momenta.

Let $\alg{A}$ be a $C^{*}$-algebra and let $(\pi
_{\rho},\hil{H}_{\rho},x_{\rho})$ be the GNS triple for $\alg{A}$
induced by the pure state $\rho$.  Once more, let $\alg{R}$ be a
family of mutually commuting observables drawn from $\alg{A}$.  Now,
$\pi _{\rho}(\alg{R})$ is a mutually commuting family of observables
in $\alg{L}(\hil{H}_{\rho})$.  Thus, we may apply
Corollary~\ref{unique}.(ii) to conclude that
$\beable{\pi_{\rho}(\alg{R})}{x_{\rho}}\equiv
\alg{L}(\hil{S}^{\perp})\oplus \vn{\pi _{\rho}(\alg{R})}P_{\hil{S}}$
is the unique maximal (in $\alg{L}(\hil{H}_{\rho})$) $\pi
_{\rho}(\alg{R})$-beable algebra for $\omega _{x_{\rho}}$.  (In this
case, $\hil{S}$ is the smallest subspace of $\hil{H}_{\rho}$ such that
$x_{\rho}\in \hil{S}$ and $\pi _{\rho}(\alg{R})$ leaves $\hil{S}$
invariant; i.e. $\hil{S}=[\pi _{\rho}(\alg{R})''x_{\rho}]$.)  We now
verify that the inverse image of $\alg{B}(\pi
_{\rho}(\alg{R}),x_{\rho})$ under $\pi _{\rho}$ is $\alg{R}$-beable
for $\rho$.

\begin{notation}  We define $\alg{B}(\alg{R},\rho)\equiv
  \pi _{\rho}^{-1}[\alg{B}(\pi_{\rho}(\alg{R}),x_\rho )]$.  This is
  meant to extend our earlier (concrete) notation $\alg{B}(\alg{R},v)$
  since, when $\alg{A}=\alg{L}(\hil{H})$ and $\rho$ is induced by a
  unit vector $v\in\hil{H}$, $\alg{L}(\hil{H})$ and
  $\alg{L}(\hil{H}_{\rho})$ are unitarily equivalent, from which it
  follows that $\alg{B}(\alg{R},\rho)=\alg{B}(\alg{R},v)$.
\end{notation}

\begin{prop} Let $\rho$ be a pure state of $\alg{A}$ and let $\alg{R}$
  be a mutually commuting family of observables in $\alg{A}$.  Then,
  $\alg{B}(\alg{R},\rho)$ is $\alg{R}$-beable for $\rho$.
  \label{any-pure} \end{prop}
\begin{proof}  Clearly,
  $\alg{B}(\alg{R},\rho)$ is a $C^{*}$-algebra, since it is the
  inverse image under $\pi_{\rho}$ of a $C^{*}$-algebra.  Furthermore,
  ($\alg{R}$-Priv) follows by the construction of
  $\alg{B}(\alg{R},\rho)$.
  
  (Beable) Let $\hil{T}\equiv [\pi _{\rho
    }(\alg{B}(\alg{R},\rho))x_{\rho}]$.  By
  Prop.~\ref{beable_equiv}.(iv), it will be sufficient to show that
  $\pi _{\rho}(\alg{B}(\alg{R},\rho))P_{\hil{T}}$ is abelian.
  Clearly, $\hil{T}$ is the smallest subspace (of $\hil{H}_{\rho}$)
  that contains $x_{\rho}$ and that is invariant under $\pi
  _{\rho}(\alg{B}(\alg{R},\rho))$.  However, $\hil{S}$ contains
  $x_{\rho}$ by construction, and $\hil{S}$ is invariant under $\pi
  _{\rho}(\alg{B}(\alg{R},\rho))$ (since $\pi _{\rho}$ maps
  $\alg{B}(\alg{R},\rho)$ into $\alg{L}(\hil{S}^{\perp})\oplus
  \alg{L}(\hil{S})$).  Thus $\hil{T}\subseteq \hil{S}$.  Conversely,
  $\pi _{\rho}(\alg{R})$ leaves $\hil{T}$ invariant, since $\pi
  _{\rho}(\alg{R})\subseteq \pi _{\rho}(\alg{B}(\alg{R},\rho))$.
  Therefore $\hil{S}=\hil{T}$ and $\pi
  _{\rho}(\alg{B}(\alg{R},\rho))P_{\hil{T}}=\pi _{\rho}
  (\alg{B}(\alg{R},\rho))P_{\hil{S}}\subseteq \vn{\pi
    _{\rho}(\alg{R})}P_{\hil{S}}$.  The conclusion then follows by
  noting that $\vn{\pi _{\rho}(\alg{R})}P_{\hil{S}}$ is abelian (since
  $\alg{R}$ is a mutually commuting family of operators).
     
  (Def) Let $U$ be a unitary element of $\alg{A}$ such that $U\in
  \alg{R}'$ and $\rho _{U}=\rho$.  In this case (i.e. where $\rho$ is
  pure), we can actually prove the stronger result that $U\in
  \alg{B}(\alg{R},\rho)$, from which it follows immediately that
  $U\alg{B}(\alg{R},\rho)U^{*} =\alg{B}(\alg{R},\rho)$.
  
  We show first that $x_{\rho}$ is an eigenvector of $\pi _{\rho}(U)$.
  For this, let $x\equiv x_{\rho}$ and let $y\equiv
  \pi_{\rho}(U)x_{\rho}$.  Since $\rho$ is pure, the representation
  $(\pi _{\rho},\hil{H}_{\rho})$ of $\alg{A}$ is
  irreducible~\cite[Thm. 10.2.3]{kad}.  Thus,
  $\pi_{\rho}(\alg{A})^{-}=\alg{L}(\hil{H}_{\rho})$.  In particular,
  there is a net $(\pi_{\rho}(H_{a}))\subseteq \pi _{\rho}(\alg{A})$
  such that $\wlim_{a}\pi_{\rho}(H_{a})=P_{x}$.  However, $\langle \pi
  _{\rho}(H_{a})x,x\rangle =\rho (H_{a})=\rho _{U}(H_{a})=\langle \pi
  _{\rho}(H_{a})y,y\rangle $, for all $a$.  Since $\omega _{x}$ and
  $\omega _{y}$ are WOT-continuous, it follows that \begin{equation}
    1\:=\:\langle P_{x}x,x\rangle \:=\:\langle P_{x}y,y\rangle
    \:=\:\abs{\langle x,y\rangle }^{2}.\end{equation} Hence,
  $\abs{\langle x,y\rangle }^{2}=\norm{x}\cdot \norm{y}$, and by the
  Cauchy-Schwarz inequality, $y=cx$ for some $c\in \mathbb{C}^{1}$,
  which is what we wanted to show.
  
  Now, since $x_{\rho}$ is an eigenvector of $\pi _{\rho}(U)$, it
  follows that $\pi _{\rho}(U)x_{\rho}\in \hil{S}$.  Moreover, since
  $U\in \alg{R}'$, it follows that $[\pi _{\rho}(U),\pi
  _{\rho}(\alg{R})]=\{ 0\}$.  Thus, by Prop.~\ref{the_test}.(i), $\pi
  _{\rho}(U)\in \alg{B}(\pi _{\rho}(\alg{R}),x_{\rho})$ and
  $U\in\alg{B}(\alg{R},\rho)$.
\end{proof}

\begin{open} Let $\hil{H}$ be a Hilbert space, and let $\rho$ be a state of $\bh$.  
\begin{enumerate} \item  When $\rho$ is singular: Do all its maximal 
  $\alg{R}$-beable algebras contain $\alg{B}(\alg{R},\rho )$?  Is
  $\alg{B}(\alg{R},\rho)$ itself maximal?  Is it unique?
\item When $\rho$ is not normal (pure or mixed): Classify the maximal
  $\alg{R}$-beable algebras for $\rho$ along the lines of
  Theorem~\ref{rbeable}.
\item When $\rho$ is not a vector state: Give necessary and sufficient
  conditions for there to be a unique maximal $\alg{R}$-beable algebra
  for $\rho$~(cf.~Corollary~\ref{unique}).  \end{enumerate}
\end{open}

In our final section, we reconstruct Bohr's reply to the original EPR
argument, in terms of maximal $\alg{R}$-beable algebras by employing
$\alg{B}(\alg{R},\rho )$, which is well-defined when $\rho$ is taken
to be the singular EPR state.  Should the answer to the second and
third questions in (i) above prove negative, the results of our
reconstruction will still remain valid if the first question can be
answered positively.

\subsection{EPR Correlations: Position/Momentum Case} \label{orig-epr}
We begin by defining the EPR state $\rho$.  Let $\hil{G}\equiv
L_{2}(\mathbb{R})$, and let $Q$ be the unbounded, self-adjoint
position operator on $\hil{G}$ defined by $Q\psi =x\psi$, where
$\hil{D}(Q)$ consists of those functions $\psi \in L_{2}(\mathbb{R})$
such that $x\psi \in L_{2}(\mathbb{R})$.  Let $T$ be the
$L_{2}(\mathbb{R})$ Fourier transform, a unitary operator on
$\hil{G}$~(cf.~\cite[Thm.  3.2.31]{kad}).  Let $D$ be the unbounded,
self-adjoint operator on $\hil{G}$ defined to be $T^{-1}QT$ on domain
$T^{-1}(\hil{D}(Q))$~(cf.~\cite[Exercise 5.7.49]{kad}).  One may show,
then, that $D\psi =i(d\psi /dx)$ when $\psi \in \hil{D}(D)$ is
differentiable, i.e., $D$ is the momentum operator.

Since $Q$ is affiliated with the abelian von Neumann algebra $\vn{Q}$,
$Q$ is represented by a self-adjoint function $\phi (Q)$ on
$\states{S}_{0}$ (the space of pure states of $\vn{Q}$).  Since
$\spectrum{Q}=\mathbb{R}$ and the range of $\phi (Q)$ is equal to
$\spectrum{Q}$, we are guaranteed that for each $\ell \in \mathbb{R}$,
there is a (necessarily singular) pure state $\alpha\in
\states{S}_{0}$ such that $\phi (Q)(\alpha)=\ell$.  We may apply the
same procedure for $D$ in order to obtain a pure state $\beta$ of
$\vn{D}$.  In keeping with the original EPR argument, we shall choose
$\beta$ (as we may) such that $\phi (D)(\beta )=0$.

Now, we may think of $\vn{Q}$ and $\vn{D}$ as acting on two different
copies $\hil{G}_{1}$ and $\hil{G}_{2}$ of $\hil{G}$.  In this case, we
may form the $C^{*}$-tensor product $\vn{Q}\otimes \vn{D}$, which acts
on $\hil{G}_{1}\otimes \hil{G}_{2}$~\cite[Section 11.1]{kad}.  It then
follows that there is a \emph{unique} pure state $\alpha \otimes
\beta$ on the $\vn{Q}\otimes \vn{D}$~\cite[Prop.  11.1.1]{kad}.
Moreover, since $\alpha \otimes \beta$ is pure, we may extend it to a
pure state $\omega$ of $\alg{L}(\hil{G}_{1}\otimes \hil{G}_{2})$.
(Note, however, that there is no guarantee of the uniqueness of our
choice of $\omega$.  In particular, arbitrariness entered into our
choice of $\alpha$ and $\beta$ as well as into our extension of
$\alpha \otimes \beta$ to $\alg{L}(\hil{G}_{1}\otimes \hil{G}_{2})$.)

\begin{notation}  Let $\hil{H}\equiv \hil{G}_{1}\otimes
  \hil{G}_{2}\simeq L_{2}(\mathbb{R}^{2})$.  Let $Q_{1}$ be the
  unbounded operator $Q\otimes I$ acting on $\hil{H}$.  Let $Q_{2}$ be
  the operator $I\otimes Q$.  Define $D_{1}$ and $D_{2}$ similarly.
\end{notation}

For $\theta \in \mathbb{R}$, let $R_{-\theta}$ be the rotation of
$\mathbb{R}^{2}$ through $-\theta$.  Let $U(\theta )$ be the unitary
rotation operator on $\hil{H}$ defined by $U(\theta )\psi \equiv \psi
\circ R_{-\theta }$.  If we let $X\equiv U(\theta )^{-1}Q_{1}U(\theta
)$ and $P\equiv U(\theta )^{-1}D_{2}U(\theta )$, it follows that
\begin{equation} X=Q_{1}\cos \theta\,\widehat{-}\,Q_{2} \sin \theta
\qquad \qquad P=D_{1}\sin \theta\,\widehat{+}\,D_{2}\cos \theta .
\end{equation} (That is, these pairs of unbounded operators have the same domain
and agree on this domain.  Cf. Bohr~\cite[p. 696(note)]{bohr}.)  With
this in mind, we define the singular state $\rho$ of $\bh$ by $\rho
\equiv \omega _{U(\theta )}$.  One can then show that for any $f\in
\mathcal{B}(\mathbb{R})$,
\begin{equation}
\rho (f(X))=\omega (f(Q_{1})) \qquad \qquad \rho (f(P))=\omega (f(D_{2}))
.\end{equation} This makes precise the sense in which the behavior
of $\rho$ relative to $X$ and $P$ is identical to the behavior of
$\omega$ relative to $Q_{1}$ and $D_{2}$.  In particular, it
follows that $\rho \,|_{\vn{X}}$ and $\rho \,|_{\vn{P}}$ are dispersion-free and
are, respectively, well-defined for $X$ and $P$.  Moreover, one can show that
$\phi (X)(\rho _{1})=\ell$ and $\phi (P)(\rho _{2})=0$, where $\rho_{1}=\rho \,|_{\vn{X}}$ and $\rho _{2}=\rho \,|_{\vn{P}}$.  In what
follows we will fix $\theta =\pi /4$.  However, instead of letting 
$X\equiv 2^{-1/2}(Q_{1}\widehat{-}Q_{2})$, we let
$X\equiv Q_{1}\widehat{-}Q_{2}$, the relative position of two 
particles moving in one-dimension; and, similarly, we let $P\equiv
P_{1}\widehat{+}P_{2}$, their total momentum.
Since $\rho$ assigns a definite (nonzero) relative position to the 
particles, and assigns them a definite (zero) total momentum, knowing the value 
of $Q_{1}$ in state $\rho$ allows one to predict with certainty the value 
of $Q_{2}$, and similarly for $D_{1}$ and $D_{2}$.  Thus we have 
the  conditions employed by EPR, in conjunction with their reality 
criterion, to argue for the simultaneous determinacy of $Q_{2}$ and 
$D_{2}$.

We pause to note a technical difficulty---a feature of the EPR state
$\rho$, not present in the singlet spin state version---that EPR do
not address.  Since $[Q_{2},P]=[D_{2},X]=i\hbar I$, Corollary
\ref{cor_cool} dictates that neither $Q_{2}$ nor $D_{2}$ has beable
status for $\rho$---or for any state obtained from $\rho$ after a
measurement on particle $1$ is performed.  Thus any argument which
purports to establish the existence of simultaneous definite values in
state $\rho$ for $Q_{2}$ and $D_{2}$ is necessarily suspect.  In fact,
since $[Q_{1},P]=-[D_{1},X]=i\hbar I$, Corollary \ref{cor_cool} also
dictates that the probability of obtaining a value in any finite
interval of the real line for $Q_{1}$ or $D_{1}$ is always zero in the
state $\rho$.  This blocks the use of EPR's reality criterion, which
first requires that either $Q_{1}$ or $D_{1}$ is measured and a finite
value obtained.  However, a natural way to overcome this obstacle is
simply to understand EPR as setting out to establish that all
\emph{bounded} Borel functions of both $Q_{2}$ and $D_{2}$ have
simultaneous reality in the state $\rho$.

Unfortunately, we have been unable to confirm that $\rho$ must dictate
strict correlations between \emph{arbitrary} Borel functions of
$Q_{1}$ and $Q_{2}$ (or $D_{1}$ and $D_{2}$).
\begin{open} Let $\hil{H}\equiv L_{2}(\mathbb{R}^{2})$, and let 
  $E_{1},E_{2},F_{1}$ and $F_{2}$ denote, respectively, the
  projection-valued measures for $Q_{1},Q_{2},D_{1}$ and $D_{2}$.  For
  any pure state $\omega$ of $\bh$, we say that $\omega$ is
  \emph{completely EPR-correlated} just in case $\omega
  (E_{2}(\reals{C}-\ell )E_{1}(\reals{C}))=\omega (E_{1}(\reals{C}))$
  and $\omega (F_{1}(\reals{C})F_{2}(\reals{C}))=\omega
  (F_{1}(\reals{C}))$ for all Borel subsets $\reals{C}$ of
  $\mathbb{R}$.
\begin{enumerate} 
\item Are there completely EPR-correlated states?
\item Must $\rho$ (as defined above) be completely EPR-correlated?
\end{enumerate} 
\end{open}
\noindent On the other hand, we shall shortly establish
(Lemma \ref{correlated} below) that $\rho$ strictly correlates the
bounded uniformly continuous (BUC) functions of $Q_{1}$ with those of
$Q_{2}$, and the BUC functions of $D_{1}$ with $D_{2}$.  Let
$\cstar{Q_{i}}$ denote the $C^{*}$-algebra of all BUC functions of
$Q_{i}$, and similarly for $\cstar{D_{i}}$.  Then we shall take as the
object of the EPR argument the establishment (at a minimum) of the
simultaneous reality of $\cstar{Q_{2}}$ and $\cstar{D_{2}}$ in $\rho$.
Should the answer to (ii) above be positive, EPR's reality criterion
would entitle them to substitute $\vn{Q_{2}}$ and $\vn{D_{2}}$ for
$\cstar{Q_{2}}$ and $\cstar{D_{2}}$; but, then the same substitution
would apply to Bohr's reply.  Note also that, since $\rho$ is not
ultraweakly continuous, such a substitution is not automatically
warranted.

Let $\vn{Q_{1},Q_{2}}$ be the abelian von Neumann algebra generated by
$Q_{1}$ and $Q_{2}$.  (The reader should note that everything we
subsequently establish about $Q_{1}$ and $Q_{2}$ in the state $\rho$
follows, by symmetry, for $D_{1}$ and $D_{2}$ as well.)  Since
$\vn{Q_{1},Q_{2}}$ is abelian, we may represent it as the space of
continuous functions on the set $\states{S}$ of pure states of
$\vn{Q_{1},Q_{2}}$.  Moreover, $\rho |_{\vn{Q_{1},Q_{2}}}$ may be
represented as a probability measure $\mu _{\rho}$ on $\states{S}$.

\begin{remark} Let $R$ be a (possibly unbounded) self-adjoint operator.  Recall
  that for any $f\in \mathcal{B}(\mathbb{R})$, the operator $f(R)$ is
  canonically constructed by employing the representation of $R$ as a
  function $\phi _{0}(R)$ in the space $\mathcal{N}(\states{S}_{0})$
  of unbounded functions on the set $\states{S}_{0}$ of pure states of
  $\vn{R}$.  Suppose that $\alg{V}$ is another abelian von Neumann
  algebra such that $R\,\eta\,\alg{V}$, and let $\states{S}_{1}$ be
  the set of pure states of $\alg{V}$.  Then, $R$ is represented by a
  function $\phi _{1}(R)$ in $\mathcal{N}(\states{S}_{1})$ In such a
  case, we may mimic the canonical construction in order to obtain
  ``functions'' of $R$ in $\alg{V}$.  Fortunately, we are guaranteed
  that, whether we perform the construction relative to $\vn{R}$ or
  relative to $\alg{V}$, there can be no ambiguity concerning the
  resulting operator $f(R)$~\cite[Remark 5.6.28]{kad}.  This fact will
  be important for what follows, since we will be concerned with
  functional relationships between $X,Q_{1}$ and $Q_{2}$.  In this
  case, all three operators are affiliated with the abelian von
  Neumann algebra $\vn{Q_{1},Q_{2}}$.  \end{remark}

In an abuse of notation (which will be justified in what follows) let
$\widetilde{g}$ denote a Borel subset of $\states{S}$.  Then, we know
that there is a \emph{unique} clopen set $h$ such that $h\triangle
\widetilde{g}$ is meager, where $h\triangle
\widetilde{g}=(h-\widetilde{g})\cup (\widetilde{g}-h)$.  Let $f$ be
another clopen set such that $\widetilde{g}\subseteq f$.  Then, noting
that $h\cap f$ is also clopen, it follows from an elementary
set-theoretic argument that $h\subseteq f$.  Using this fact, we may
establish the following lemma:

\begin{lemma} Let $\states{L} \equiv \{ \omega \in \states{S}:\phi
  (X)(\omega )=\ell \}$.  Then, $\mu _{\rho}(\states{L} )=1$.
\label{epr-value} \end{lemma}

\begin{proof} Fix $n\in \mathbb{N}$ and let $f$ be the characteristic
  function of the clopen set \begin{equation} \states{S}_{n}\equiv
    \bigl[ \phi (X)^{-1}(\ell -n^{-1},\ell +n^{-1})\bigr] ^{-}
    .\end{equation} By the construction of $\rho$, we have $\rho
  (E(\reals{C}_{n}))=1$ where $E$ is the spectral-measure for $X$ and
  $\reals{C}_{n}\equiv (\ell -n^{-1},\ell +n^{-1} )\subseteq
  \mathbb{R}$ (cf. Lemma \ref{yes}.(ii)).  In other words, $\int h\,
  d\mu _{\rho}=1$, where $h=\phi (E(\reals{C}_{n}))$.  Recall that $h$
  is defined to be the unique closest continuous function to
  $\widetilde{g}$, where $\widetilde{g}(\omega )=1$ if $\phi
  (X)(\omega )\in (\ell -n^{-1},\ell +n^{-1})$ and
  $\widetilde{g}(\omega )=0$ otherwise.  However, $\widetilde{g}\leq
  h$, for if $\widetilde{g}(\omega )=1$, then $\omega \in \phi
  (X)^{-1}(\ell -n^{-1},\ell +n^{-1})$.  Now applying the
  considerations prior to this lemma (identifying sets with their
  characteristic functions), we have $h\leq f$ and $\int f\,d\mu
  _{\rho}=1$.  That is, $\mu _{\rho}(\states{S}_{n})=1$ for all $n\in
  \mathbb{N}$.  Moreover, since $\states{S}_{n}\supseteq
  \states{S}_{n+1}$ for all $n$, $\mu _{\rho}(\cap
  \states{S}_{n})=\lim _{n}\mu _{\rho }(\states{S}_{n})=1$.  Since
  $\cap _{n\in \mathbb{N}}\states{S}_{n}\subseteq \states{L}$, it
  follows that $\mu _{\rho}(\states{L} )=1$. \end{proof}

\begin{lemma}  Let $\states{Z}$ and $\states{Z}'$ be the closed,
  nowhere dense subsets of $\states{S}$ at which $\phi (Q_{1})$ and
  $\phi (Q_{2})$, respectively, are not defined.  Then $\states{L}$ is
  the disjoint union of $\states{L}\cap (\states{Z}\cap \states{Z}')$
  and $\states{L}\backslash (\states{Z}\cup \states{Z}')$.
\label{same}
\end{lemma}

\begin{proof}  Let $\states{Z}''$ be the set of points in $\states{S}$ at which
  $\phi (Q_{1})\widehat{-}\phi (Q_{2})$ is not defined, and suppose
  that $\omega \in \states{L}\cap \states{Z}\,\subseteq
  (\states{S}\backslash \states{Z}'')\cap \states{Z}$.  Since
  $\states{Z}\cup \states{Z}'\cup \states{Z}''$ is closed and nowhere
  dense, there is a net $(\tau _{a})\subseteq \states{S}\backslash
  (\states{Z}\cup \states{Z}'\cup \states{Z}'')$ such that $\tau
  _{a}\!\rightarrow \!\omega$.  Using the fact that $\phi
  (Q_{1}\widehat{-}Q_{2})(\tau _{a})=\phi (Q_{1})(\tau _{a})-\phi
  (Q_{2})(\tau _{a})$ for each $\tau _{a}$ (since $\tau _{a}\in
  \states{S}\backslash (\states{Z}\cup \states{Z}')$), and the fact
  that $\lim _{a}\abs{\phi (Q_{1})(\tau _{a})}=\infty$, it follows
  that $\lim _{a}\abs{\phi (Q_{2})(\tau _{a})}=\infty \,$, and thus
  $\omega \in \states{Z}'$.  A similar argument shows that if $\omega
  \in \states{L}\cap \states{Z}'$, then $\omega \in \states{Z}$.
\end{proof}

\begin{lemma} For any $A\in \cstar{Q_{2}}$, there is an $A'\in
  \cstar{Q_{1}}$ such that $\omega (A')=\omega (A)$ for all $\omega
  \in \states{L}$ (and $A'$ may be chosen so that
  $\spectrum{A'}=\spectrum{A}$).  \label{correlated} \end{lemma}

\begin{proof}  Let $A\in \cstar{Q_{2}}$.  Then, $A=f(Q_{2})$, for some
  BUC function $f$ on $\mathbb{R}$.  Let $A'\equiv g(Q_{1})$, where
  $g(x)=f(x-\ell ),\;(x\in \mathbb{R})$, and let $h_{1}$ and $h_{2}$
  be the unique continuous functions on $\states{S}$ corresponding,
  respectively, to $\widetilde{f}$ and $\widetilde{g}$.  (Clearly,
  $h_{1}$ and $h_{2}$ have identical range and it follows
  from~\cite[Prop.~5.6.20]{kad} that $\spectrum{A}=\spectrum{A'}$.)
  
  We now show that that
  $\widetilde{f}\,|_{\states{L}}=\widetilde{g}\,|_{\states{L}}$.  For
  this, let $\omega \in \states{L}$.  (Case 1a) Suppose that $\omega
  \in \states{L}- (\states{Z}\cup \states{Z}')$.  Then $\ell=\phi
  (Q_{1})\widehat{-}\phi (Q_{2})(\omega )=\phi (Q_{1})(\omega )-\phi
  (Q_{2})(\omega )$ and $\phi (Q_{1})(\omega )-\ell =\phi
  (Q_{2})(\omega )$.  Therefore,
\begin{equation} \widetilde{g}(\omega )\:=\:g(\phi (Q_{1})(\omega ))\:=\:f(\phi
(Q_{1})(\omega )-\ell )\:=\:f(\phi (Q_{2})(\omega ))\:=\:\widetilde{f}(\omega ).
\end{equation}

(Case 1b) Suppose that $\omega \in \states{L}\cap(\states{Z}\cap
\states{Z}')$.  Then, by definition, $\widetilde{f}(\omega
)=\widetilde{g}(\omega )=0$, since $\phi (Q_{1})$ and $\phi (Q_{2})$
are not defined at $\omega $.

We now show that $h_{1}\,|_{\states{L}}=h_{2}\,|_{\states{L}}$.  (Case
2a) Suppose that $\omega \in \states{L}-(\states{Z}\cup \states{Z}')$.
By Case~1a, it will be sufficient to show that $h_{1}(\omega
)=\widetilde{f}(\omega )$ and $h_{2}(\omega )=\widetilde{g}(\omega )$.
In order to establish this, note that $\widetilde{f}$ and
$\widetilde{g}$ are continuous on $\states{S}-(\states{Z}\cup
\states{Z}')$ (since each is the composition of two continuous
functions).  Moreover, by definition, $\widetilde{f}$ may not disagree
with the continuous function $h_{1}$ on any open set (in
$\states{S}$), and $\widetilde{g}$ may not disagree with the
continuous function $h_{2}$ on any open set (in $\states{S}$).
Therefore, $h_{1}\,|_{\states{S}-(\states{Z}\cup
  \states{Z}')}=\widetilde{f}\,|_{\states{S}-(\states{Z}\cup
  \states{Z}')}$ and $h_{2}\,|_{\states{S}-(\states{Z}\cup
  \states{Z}')}=\widetilde{g}\,|_{\states{S}-(\states{Z}\cup
  \states{Z}')}$.

(Case 2b) Suppose that $\omega \in \states{L}\cap (\states{Z}\cap
\states{Z}')$.  Since $\states{Z}\cup \states{Z}'$ is nowhere dense,
there is a net $(\tau _{a})\subseteq \states{S}\backslash
(\states{Z}\cup \states{Z}')$ such that $\tau_{a}\!\rightarrow\!\omega
$.  For each $a$,
\begin{align} \bigl|\,h_{1}(\omega )-h_{2}(\omega )\,\bigr|&=\bigl|\,h_{1}(\omega
  )-h_{1}(\tau _{a})+h_{1}(\tau _{a}) -h_{2}(\tau _{a})+h_{2}(\tau
  _{a})-h_{2}(\omega )\,\bigr| \\
  &\leq \bigl|\,h_{1}(\omega )-h_{1}(\tau
  _{a})\,\bigr|\;+\;\bigl|\,h_{1}(\tau _{a})-h_{2}(\tau
  _{a})\,\bigr|\;+\;\bigl|\,h_{2}(\tau _{a})-h_{2}(\omega )\,\bigr|
\label{approx}.\end{align}   If we take the limit over $a$ of the RHS of~(\ref{approx}), the first and third terms go to zero since
$h_{1}$ and $h_{2}$ are continuous.  Thus, $\abs{h_{1}(\omega
  )-h_{2}(\omega )}\leq \lim _{a}\abs{h_{1}(\tau _{a})-h_{2}(\tau
  _{a})}$.

If we let $x_{a}\equiv \phi (Q_{2})(\tau _{a} )$ and $y_{a}\equiv \phi
(Q_{1})(\tau _{a})-\ell$, then it follows from the continuity of $f$
that $h_{1}(\tau _{a})=f(x_{a})$ and $h_{2}(\tau
_{a})=f(y_{a})$~(see~Case~2a).  Thus, $\abs{h_{1}(\omega
  )-h_{2}(\omega )}\leq \lim_{a}\abs{f(x_{a})-f(y_{a})}$.  Now, using
the fact that $\phi (Q_{1})\widehat{-}\phi (Q_{2})$ is continuous at
$\omega $, and $(\tau _{a})\subseteq \states{S}\backslash
(\states{Z}\cup \states{Z}')$, we have
\begin{equation} \ell \:=\:[\phi (Q_{1})\widehat{-}\phi
(Q_{2})](\omega )\:=\:\lim _{a}[\phi (Q_{1})\widehat{-}\phi
(Q_{2})](\tau _{a})\:=\:\lim _{a}[\phi (Q_{1})(\tau _{a})-\phi
(Q_{2})(\tau _{a})] ,\end{equation} and $\lim
_{a}\abs{x_{a}-y_{a}}=0$.  But this, in conjunction with the fact that
$f$ is uniformly continuous entails that
$\lim_{a}\abs{f(x_{a})-f(y_{a})}=0$.  Therefore, $h_{1}(\omega
)=h_{2}(\omega )$.
\end{proof}

\begin{remark} Lemma~\ref{correlated}, and its analogue for $D_{1}$ 
  and $D_{2}$, is necessary for EPR to be able to use $\rho$ to argue
  for the simultaneous determinacy of $\cstar{Q_{2}}$ and
  $\cstar{D_{2}}$.  The reader will note that this lemma cannot be
  straightforwardly modified for the case where $A=f(Q_{1})$ for any
  $f$ in $\mathcal{B}(\mathbb{R})$ or in $\mathcal{C}(\mathbb{R})$
  (cf. the open problem above).  On the one hand, the assumption of
  continuity is needed to show that $h_{1}(\omega )=h_{2}(\omega )$
  when $\omega \in \states{L}-(\states{Z}\cup \states{Z}')$.  On the
  other hand, the assumption of \emph{uniform} continuity is needed to
  show that $h_{1}(\omega )=h_{2}(\omega )$ when $\omega \in
  \states{L}\cap (\states{Z}\cap \states{Z}')$.  \end{remark}

Lastly, we turn to Bohr's reply.  As in our earlier analysis of the
spin version of EPR's argument, we need to consider the effect of an
ideal measurement of $Q_{1}$ that strictly correlates its values to
those of an apparatus, initially in a ground state $\omega_{0}$, with
a pointer observable $R$ satisfying $\spectrum{R}=\spectrum{Q_{1}}$.
The final post-measurement state of apparatus and particles will have
the form $(\omega_{0}\otimes\rho)_{U}$, where the unitary evolution
effecting the measurement correlation satisfies $[U,Q_{1}]=0$
(consistent with the measurement being ideal).  Observing the
registered value for any element of $\cstar{R}$, the value of the
corresponding element of $\cstar{Q_{1}}$ may then be inferred.  If we
again understand Bohr's reply in terms of selecting the appropriate
maximal $\alg{R}$-beable algebra for this measurement context, he can
be seen (modulo our remarks at the end of last section) as endorsing
the attribution of determinate values to the elements in
$\alg{B}(\cstar{R},(\omega _{0}\otimes\rho)_{U})$.  It is not
difficult to show (given the above specifications of the measurement
interaction) that the set $\alg{B}(\cstar{Q_{1}},\rho )$ coincides
with the elements of $\alg{B}(\cstar{R},(\omega _{0}\otimes\rho)_{U})$
that pertain only to the two EPR particles.  Thus, our final
proposition below establishes, in direct analogy to the spin case,
that $\alg{B}(\cstar{R},(\omega_{0}\otimes\rho)_{U})$ contains all BUC
functions of $Q_{2}$ but \emph{not} of $D_{2}$.

\begin{prop}  \ \ 
\begin{enumerate}
\item $\cstar{Q_{2}}\subseteq\alg{B}(\cstar{Q_{1}},\rho )$.
\item $\cstar{D_{2}}\not\subseteq\alg{B}(\cstar{Q_{1}},\rho )$.
\end{enumerate}
\label{bohr}  \end{prop}

\begin{proof} (i) Since $\cstar{Q_{2}}$ is a $C^{*}$-algebra, it will be sufficient to
  show that for every unitary element $A\in \cstar{Q_{2}}$, $A\in
  \alg{B}(\cstar{Q_{1}},\rho)$.  Moreover, since $\pi _{\rho}(A)$
  commutes with all elements in $\pi _{\rho}(\cstar{Q_{1}})$, the
  result would follow from Prop.~\ref{the_test}.(i) if we could show
  that $\pi _{\rho}(A)x_{\rho}\in \hil{S}$.  We proceed to show this.

  From Lemma~\ref{correlated}, there is a unitary $A'\in
  \cstar{Q_{1}}$ such that $\omega (A)=\omega (A')$, for all $\omega
  \in \states{L}$.  Since each $\omega \in \states{L}$ is
  dispersion-free, it follows that $\omega
  ((A')^{*}A)=\overline{\omega (A')}\omega (A)=\abs{\omega
    (A)}^{2}=1$.  Thus, \begin{equation} \rho ((A')^{*}A)=\int
    _{\states{S}}\omega _{s}((A')^{*}A)d\mu _{\rho}(s)\:=\int
    _{\states{L}}\omega _{s}((A')^{*}A)d\mu _{\rho}(s)\:=\:\mu
    _{\rho}(\states{L} ) \:=\: 1, \label{unitaries}
\end{equation}  where we have used Lemma~\ref{epr-value} in the second
and final equalities.  From~(\ref{unitaries}) it follows that, in the
GNS representation (of $\bh$) for $\rho$, $\langle \pi
_{\rho}((A')^{*}A)x_{\rho},x_{\rho} \rangle =1$.  Hence, we may use
the fact that $\pi _{\rho}$ is a $*$-homomorphism in combination with
the Cauchy-Schwarz inequality to conclude that $\pi
_{\rho}(A)x_{\rho}=c\pi _{\rho}(A')x_{\rho}$, for some $c\in
\mathbb{C}^{1}$.  In particular, $\pi _{\rho}(A)x_{\rho}\in [\pi
_{\rho}(\cstar{Q_{1}})x_{\rho}]=\hil{S}$, as we wished to show.

(ii) Since $\beable{C^{*}(Q_{1})}{\rho}$ is beable for $\rho$, it has
a dispersion-free state $\omega$.  We show that this entails that
$W_{s}\equiv e^{isD_{2}}\not\in \beable{C^{*}(Q_{1})}{\rho}$ for all
$s\neq 0$.  In order to see this, note first that $U_{t}\equiv
e^{itQ_{2}}\in C^{*}(Q_{2})\subseteq \beable{C^{*}(Q_{1})}{\rho}$, for
all $t\in \mathbb{R}$, since $e^{itx}$ is uniformly continuous on
$\mathbb{R}$.  Suppose, for reductio ad absurdem, that $W_{s}\in
\beable{C^{*}(Q_{1})}{\rho}$ for some $s\neq 0$.  Since $Q_{2}$ and
$D_{2}$ satisfy the Weyl-form of the CCR we have
$U_{t}W_{s}=e^{ist}W_{s}U_{t}$ for all $t\in \mathbb{R}$, and $\omega
(U_{t}W_{s})=e^{ist}\omega (W_{s}U_{t})$ for all $t\in \mathbb{R}$.
Fix $t\in \mathbb{R}$ such that $st\neq n\pi$ for any $n\in
2\mathbb{Z}$.  Since $\omega$ is dispersion-free on $U_{t}$, it
follows that $\omega (U_{t})\omega (W_{s})=e^{ist}\omega (W_{s})\omega
(U_{t})$.  Moreover, $\omega (U_{t})\neq 0$ and $\omega (W_{s})\neq 0$
since $\omega$ must assign each unitary operator a value in its
spectrum~(Prop.~\ref{mult_dfree}.(ii)).  Thus, we have $e^{ist}=1$
contrary to our assumption that $st\neq n\pi$ for any $n\in
2\mathbb{Z}$.  Therefore $W_{s}\not\in \beable{C^{*}(Q_{1})}{\rho}$
when $s\neq 0$.  \end{proof}

By symmetry of reasoning, if we suppose that the BUC functions of
$D_{1}$ are actually measured in the original EPR experiment, instead
of those of $Q_{1}$, it will become legitimate to regard all BUC
functions of $D_{2}$, but \emph{not} of $Q_{2}$, as having determinate
values.  And, as in the spin case, one has no grounds within the
Copenhagen interpretation (so reconstructed) for asserting that both
$\cstar{Q_{2}}$ and $\cstar{D_{2}}$ are determinate in state $\rho$
relative to any fixed measurement context for particle~$1$.
\vspace{4em}

\emph{Acknowledgements:} The authors wish to thank R.W. Heath
(Pittsburgh), N. P. Landsman (Amsterdam), N.C. Phillips (Oregon), and
J.E. Roberts (Rome) for helpful correspondence.

\newpage \bibliographystyle{amsplain} \bibliography{jlblast}

\end{document}